\documentclass{aa}  

\usepackage{graphicx}
\usepackage{txfonts}
\usepackage{xcolor}
\usepackage{hyperref} 
\begin{document}

    \title{Magnetic Fields or Overstable Convective Modes in HR~7495:}

   \subtitle{ Exploring the Underlying Causes of the Spike in the 'Hump \& Spike' Features\thanks{Based on observations made with the Nordic Optical Telescope, owned in collaboration by the University of Turku and Aarhus University, and operated jointly by Aarhus University, the University of Turku and the University of Oslo, representing Denmark, Finland and Norway, the University of Iceland and Stockholm University at the Observatorio del Roque de los Muchachos, La Palma, Spain, of the Instituto de Astrofisica de Canarias.}}

   \author{V. Antoci
          \inst{1}
          \and
          M. Cantiello\inst{2,3} 
           \and 
          V. Khalack\inst{4}
           \and 
          A. Henriksen\inst{1}
           \and
           H. Saio\inst{5}
           \and 
          T. R. White\inst{6} 
           \and
          L. Buchhave\inst{1} 
          }

  \institute{DTU Space, Technical University of Denmark, Elektrovej 327, Kgs. Lyngby, 2800, Denmark\\
              \email{antoci@space.dtu.dk}   
        \and
             Center for Computational Astrophysics, Flatiron Institute, 162 5th Avenue, New York, NY 10010, USA
        \and
            Department of Astrophysical Sciences, Princeton University, Princeton, NJ 08544, USA
        \and
            D\'epartement de Physique et d'Astronomie, Universit\'e de Moncton, Moncton, NB E1A 3E9, Canada
           \and
             Astronomical Institute, Graduate School of Science, Tohoku University, Sendai 980-8578, Japan
        \and
            Sydney Informatics Hub, Core Research Facilities, University of Sydney, NSW 2006, Australia    
             }

   \date{}

% \abstract{}{}{}{}{} 
% 5 {} token are mandatory
 
\abstract
  {More than 200 A- and F-type stars observed with \textit{Kepler} exhibit a distinctive `hump \& spike' feature in their Fourier spectra. The hump is commonly interpreted as unresolved Rossby modes, while the spike has been linked to rotational modulation. This feature has led to the dubbing of these stars as `hump \& spike' stars. However, two competing interpretations exist for the spike.}
  {This study aims to determine whether the observed spike in these stars is caused by magnetic phenomena, such as spots on the stellar surface, or by Overstable Convective (OsC) modes resonantly exciting low-frequency g modes within the star's envelope.}
  {We analysed photometric data from \textit{Kepler} and TESS covering 4.5 years and four seasons, respectively, for HR~7495, the brightest `hump \& spike' star with a visual magnitude of 5.06. Additionally, radial velocity measurements and spectropolarimetric data from three different epochs were incorporated to investigate magnetic fields and surface features. Furthermore, we analysed model-based artificial light and radial velocity curves to examine the influence of OsC modes on the phase-folded light curves.} 
  {The analysis of phase-folded light curves indicates that the spike characteristics of HR~7495 align more closely with surface rotational modulation by stellar spots than with OsC modes. No significant magnetic fields were detected, which constrains the possible amplitude and geometry of the field. This is consistent with the hypothesis of a subsurface convective layer operating a dynamo, resulting in low-amplitude magnetic fields with potentially complex geometries. The variability patterns suggest multiple evolving spots. Comparing observed light and RV data with modelled OsC modes reveals a phase offset of 0.5, strongly disfavoring pulsations as the cause of the spike.}
  {While the evolutionary stage of HR~7495 does not entirely preclude the possibility of OsC modes, the observational data overwhelmingly support the stellar spots hypothesis. Our analysis, combined with previous literature, suggests that if not all A- and F-type, at least the `hump \& spike' stars harbour an undetected weak magnetic field, likely driven by a dynamo mechanism.}

\keywords{stars: oscillations (including pulsations) -- stars: activity -- stars: magnetic field -- techniques: photometric -- techniques: spectroscopic}

   \keywords{stars --
                magnetic fields --
                rotational modulation
               }

   \maketitle
%
%-------------------------------------------------------------------

\section{Introduction}\label{sec:intro}

Magnetic fields are among the most complex phenomena in astrophysics and are frequently invoked to reconcile discrepancies between theoretical models and observational data. They interact with atomic diffusion, turbulence, and rotation and notably influence angular momentum transport, critically impacting stellar evolution across various stages \citep[e.g.,][]{Maeder2003, Brun2005, Alecian2007, Landstreet2008, Featherstone2009, Keszthelyi2023, Moyano2023, Sarkar2024}. Despite their acknowledged significance, our general understanding of magnetic fields remains limited, including their strengths, geometrical configurations, origins, and, from the observational perspective, even their very existence.
In stars of masses below approximately 1.3 M$_{\odot}$, the presence of stellar magnetic fields is widely recognised \citep[e.g.,][]{Donati2009}; however, for more massive  A- and F-type stars on the main sequence, magnetic fields have been definitively detected in only about 10\% of stars. Among these, a majority are chemically peculiar Ap stars with strong, well-organised magnetic fields of the order of kilogauss. Conversely, a minority exhibit weak magnetic fields, typically less than about 100 gauss, as identified in several studies \citep{Petit2011, Neiner2017, Zwintz2020, Blazere2016, Blazere2020, Ligners2009, Neiner&Lampens2015, Thomson-Paressant2023}. From the theoretical standpoint, weak magnetic fields are predicted to potentially be ubiquitous in OBA stars due to their rapid rotation and the presence of subsurface convective regions \citep{Cantiello_hotspots_2011, Cantiello_2019_dynamo}. The observed dichotomy in magnetic field amplitude could then be due to the interplay between fossil fields and the destabilising mechanisms provided by envelope convection \citep{Jermyn:2020} and/or rotation \citep{Auriere:2007}: in this picture, the 10\% strong, well-organised magnetic fields are fossil fields with large enough amplitudes to inhibit these destabilising forces in the star. However, weak fields with complex topologies could result when these destabilising processes are able to efficiently remove weak fossil fields, replacing them with low-amplitude, dynamo-generated ones \citep{Cantiello_2019_dynamo}.\footnote{But see \citet{Braithwaite2013_failedfossile} for a different scenario able to produce weak magnetic fields starting from fossil fields in rotating stars.} 

Direct observation of weak fields through spectropolarimetry is feasible solely for very bright stars with relatively small values of $v \sin i$, suggesting the potential existence of numerous undetected surface weak magnetic fields in A and F stars. To address this detection challenge, indirect indicators of magnetic activity, such as stellar photospheric spots, have been employed. Thanks to missions like  \textit{Kepler}  \citep{Koch2010_Kepler}) and TESS  \citep[Transiting Exoplanet Survey Satellite;][]{Ricker2015}, detecting rotational modulation in hundreds of intermediate A- and F-type stars has become possible \citep[e.g.,][]{Balona2017_starspots, David-Uraz2019, Sikora2020, Trust2020, Henriksen2023a}. This modulation provides valuable insights into both strong large-scale and weaker magnetic fields.

Observations of corotating stellar spots, which induce rotational modulation, are relatively straightforward when the signal-to-noise ratio is sufficient, and the geometry is favourable, as evidenced by light curves from the \textit{Kepler} and TESS missions. However, elucidating the origins and causes of the aforementioned variability entails a more nuanced analysis. It is crucial to ascertain the presence of binary companions or potential contamination from nearby active stars, in addition to considering all plausible physical processes that could give rise to rotational modulation. Initial interpretations by \citet{Saio2018} suggested that the distinct `hump \& spike' pattern observed in the Fourier spectra of certain stars results from unresolved Rossby modes (the hump), which are mechanically excited by deviated flows due to stellar spots at intermediate to high latitudes (the spike). Subsequently, \citet{Henriksen2023a, Henriksen2023b} expanded on this work by identifying over 200 stars within the \textit{Kepler} dataset that exhibit this specific pattern, further confirming the widespread occurrence and potential implications of these phenomena. However, \citet{Lee2021, LeeSaio2020} proposed an alternate hypothesis for the spike, suggesting that overstable convective (OsC) modes present in rapidly rotating main-sequence stars could resonantly excite low-frequency g modes within the stellar envelope. In this scenario, the convective stellar core rotates marginally faster than its envelope, with the g modes manifesting observable amplitudes at the photosphere, causing the spike. In this case, it would also be rotational modulation, but the spike would correspond to the rotational frequency of the convective core.

\citet{Henriksen2023a} performed an ensemble study on nearly 200 `hump \& spike' stars, demonstrating that the spikes are more consistent with stellar spots generated by dynamo-driven magnetic fields. This is evidenced by their higher amplitudes in cooler and slightly evolved stars, aligning with theoretical predictions for dynamo activity that predict stronger magnetic activity in stars with deeper convective envelopes \citep{Cantiello_2019_dynamo}. The spike lifetimes, not exceeding a few tens of days, align with expectations for the temporal behaviour of magnetic features, while their occurrence shows no dependence on stellar mass, challenging theories that link spike frequencies with convective core rotation. This is because, in the case of OsC modes, the temporal variability of the spike would mirror the movement of convective cells inside the core, implying that the spike lifetimes would be on the convective turnover timescale. If that were the case, a correlation between spike lifetime and stellar mass should be present, which is not observed. Furthermore, harmonic analysis and phase-folded light curves revealed a non-sinusoidal pattern akin to what is expected from corotating stellar spots. Although magnetic field strength estimates reported by \citet{Henriksen2023a} were tentative, they aligned with the projections for non-Ap stars \citep{Cantiello_hotspots_2011, Cantiello_2019_dynamo}. Notably, the identification of stars potentially in a post-main-sequence phase argues against overstable convective modes due to the absence of a convective core, albeit the evolutionary phase needs to be determined from more detailed modelling. This body of evidence, complemented by recent spectropolarimetric observations of stars like $\beta$ Cas \citep{Zwintz2020}, supports the hypothesis that intermediate-mass stars can harbour dynamo-generated magnetic fields. However, detailed analyses of individual single stars are required to reach a clear conclusion.

In this article, we present in-depth analyses of the brightest `hump \& spike' star, HR~7495 (V=5.06), observed photometrically from space with \textit{Kepler} and TESS. We further collected spectroscopic data with FIES at the NOT (Nordic Optical Telescope) and performed spectropolarimetric observations with ESPaDOnS at the Canada-France-Hawaii Telescope (CFHT)\footnote{The Canada-France-Hawaii Telescope (CFHT) is operated by the National Research Council of Canada, the Institut National des Sciences de l'Univers of the Centre National de la Recherche Scientifique of France, and the University of Hawaii. The operations at the Canada-France-Hawaii Telescope are conducted with care and respect from the summit of Maunakea, which is a significant cultural and historic site.}. We compare contemporaneous radial velocity and flux measurements, analyse the shape of the phase-folded light curves, and discuss the temporal behaviour in the context of stellar spots and OsC modes. We further discuss the evolutionary stage of HR~7495 and estimate the spot size/contrast based on the non-detection of a magnetic field.

%--------------------------------------------------------------------
\section{Data }\label{sec:data}

This section details the datasets analysed to identify the spike's origin in HR~7495. We outline the sources and methodology behind our data collection.

\subsection{Photometry}\label{sec:phot}

HR~7495 (V=5.06, HD~186155, KIC~9163520, TIC~271545295, RA 19h 40m 50.18s; Dec +45$^\circ$ 31' 29.8'') is a luminous F5 II-III spectral type star within the \textit{Kepler} field. Utilising data from Gaia DR3 \citep{GaiaDR3} alongside the GSP-Phot and FLAMES modules, we have extracted the parameters listed in Table \ref{table:HR7495_params}, which also includes other astrophysical values. The GSP-Phot module (General Stellar Parametrizer from Photometry) employs low-resolution spectra from the BP/RP instruments to calculate the effective temperature, while the FLAMES module ascertains the luminosity, leveraging parallaxes, GSP-Phot inputs, and an extinction map from DR2 photometry \citep{Lallement2019}. Notably, the temperature uncertainty derived from GSP-Phot adopted an error margin of approximately 110 K, as recommended by \citep{Duerfeldt2024}.

\begin{table}[h]

\caption{Astrophysical parameters of HR~7495. The spike parameters are based on \textit{Kepler} data. }
\label{table:HR7495_params}
\centering
\begin{tabular}{l c}
\hline\hline
Stellar Parameter & Value \\
\hline
Parallax [mas] & 19.75 $\pm$ 0.07 \\
Luminosity [L$_\odot$] & 18.49$^{+0.13}_{-0.12}$ \\
T$_{\rm{eff}}$ [K] & 6610$\pm$110 \\
Radius [R$_\odot$] & 3.30$\pm$ 0.07\\
Spike frequency* [cd$^{-1}$] & $0.6770\pm0.0005$ \\
Spike period* [d] & $1.4771\pm0.0011$  \\
Spike Amplitude* [ppm]& $173.3 \pm 0.8 $ \\
Spike lifetime* [d]& $19 \pm 0.1$ \\ 
Hump width* [cd$^{-1}$]& 0.604-0.668 \\ 
$v \sin i$* [kms$^{-1}$] & $41 \pm 4 $   \\
inclination $i$* [$\deg$] & $21 \pm 2$ \\
rotational velocity $v_{\rm rot}$* [kms$^{-1}$]& $113 \pm 2$ \\

\hline
*Values from \citet{Henriksen2023a}.
\end{tabular}
\end{table}

Due to its brightness, HR~7495 was not directly observed with \textit{Kepler}. However, using the \textit{Kepler} Smear Campaign \citep{Pope2019}, long-cadence data were extracted from the entire mission, with only Quarter 0 missing. The length of the data set is 1458.5 days (Table \ref{table:data}). In the \textit{Kepler} Smear Campaign, \citet{Pope2019} utilised a method to construct light curves for stars that were otherwise too bright to be directly observed by the \textit{Kepler} telescope due to saturation. Because the \textit{Kepler} camera does not have a shutter, the light continued to fall on the CCD during read-out, resulting in a smearing of the image as the charges in each pixel were shuffled along columns. To correct for this smear, `masked' and `virtual' rows on either side of the science pixels measured the amount of light that fell on each column during read-out so that it could be subtracted. The saved `collateral' smear data could then be used to perform photometry on bright stars that dominated the flux in their columns \citep{Kolodziejczak2011, Pope2016}, as was the case with HR~7495.

The TESS data analysed in this work are the 2-min Pre-search Data Conditioning Simple Aperture Photometry (PDCSAP) light curves provided by the TESS Science Team, which are publicly available from the Mikulski Archive for Space Telescopes (MAST\footnote{https://mast.stsci.edu/portal/Mashup/Clients/Mast/Portal.html}). 
The PDCSAP are  SAP data from which long-term trends have been removed using so-called Co-trending Basis Vectors (CBVs). PDCSAP data are usually cleaner than the SAP flux and will have fewer systematic trends, which is essential when looking at periods of the order of days or longer. Although this correction may, in some cases, also remove astrophysical signals, this is not a concern in the present work, as we actively remove longer trends through bandpass filtering (see Section \ref{sec:analyses} for details).

\begin{table*}
\caption{Summary of photometric and spectroscopic observations used in the analysis of HR~7495. The data include different instruments and observing runs from Kepler, TESS, FIES@NOT, and ESPaDOnS@CFHT. Time spans are provided in days, and observing epochs are indicated where applicable. The "Contemporaneous" column marks data sets taken during overlapping time periods. We distinguish between the different years of TESS data because the datasets were analysed independently to avoid alias peaks introduced by the large gaps.}
\label{table:data}
\centering
\begin{tabular}{c c c c c c c c}        % 8 columns for consistent presentation
\hline\hline                 
Instrument & Year & Method & Time Span & Observing  & Quarters/ & Contemporaneous \\   
 &  &  &  [days] &  Epochs &Sectors &  \\   
\hline                        
\textit{Kepler} & 2009–2013 & Photometry & 1459.5 & Q1-Q17  & 17 Quarters & No \\      
TESS & 2019 & Photometry & 54 & Sectors 14, 15 & 2 Sectors & No \\  
TESS & 2021 & Photometry & 54 & Sectors 40, 41 & 2 Sectors & No \\  
TESS & 2022 & Photometry & 27 & Sector 55 & 1 Sector & Yes \\  
TESS & 2024 & Photometry & 54 & Sectors 75, 76 & 2 Sectors & No \\  
FIES@NOT & 2022 & Spectroscopy (RV) & – & 10 epochs & – & Yes \\  
ESPaDOnS@CFHT & 2023 & Spectropolarimetry & – & 3 epochs & – & No \\  
\hline     

\end{tabular}
\end{table*}

\subsection{Spectroscopy}\label{sec:spec}

We acquired contemporaneous stellar spectra alongside the TESS observations in sector 55 using the Fiber-fed Echelle Spectrograph (FIES; \citet{Telting_FIES_2014}) at the Nordic Optical Telescope (NOT; \citet{Djupvik_NOT2010}) located at the Roque de los Muchachos Observatory in La Palma, Spain. We used the higher resolution fibre with R = 67,000 and exposed HR~7495 for 80 sec. The reduction of the FIES spectra followed the methodology outlined by \citet{Buchhave2010}, involving bias subtraction, flat fielding, order tracing and extraction, and wavelength calibration. The latter utilised ThAr spectra obtained immediately before or after each scientific exposure to ensure accuracy. Radial velocities (RVs) were determined through multi-order cross-correlation \citep{Buchhave2010}, employing the spectrum with the highest signal-to-noise ratio (SNR) as a reference template (BJD=2459772.5348040, see Table \ref{table:rv_data}). The RV data were intended to be observed during the TESS observing run in 2022, which covered Sectors 54 and 55. However, there are currently no TESS data available for Sector 54 on MAST.

%--------------------------------------------------------------------

\subsection{Spectropolarimetry}\label{sec:specpol}

HR~7495 was observed with the spectropolarimeter ESPaDOnS (Echelle SpectroPolarimetric Device for Observations of Stars) at the CFHT in July and October 2023 (see Table~\ref{table:specpol}). 
Employing the deep-depletion e2v device Olapa, ESPaDOnS
acquired high-resolution (R=65,000) Stokes I \& V spectra of HR~7495 in the spectral domain from 3700\AA\, to 10000\AA\, with SNR$\geq$800. The optical characteristics of the spectrograph, as well as the instrument performances, are described in detail by \citet{Donati06}\footnote{For more details about this instrument, the reader is invited to visit {\rm www.cfht.hawaii.edu/Instruments/Spectroscopy/Espadons/}}. The dedicated software package Libre-ESpRIT \citep{Donati97} has been employed to reduce the non-polarised Stokes I and the circular polarisation Stokes V spectra. Individual normalisation was applied to all the spectra to optimise the final SNR.

\begin{figure}
   \centering
   \begin{tabular}{cc}
   \includegraphics[width=4.7cm, angle=0,trim={1.2cm 0.5cm 16cm 0.8cm}, clip]{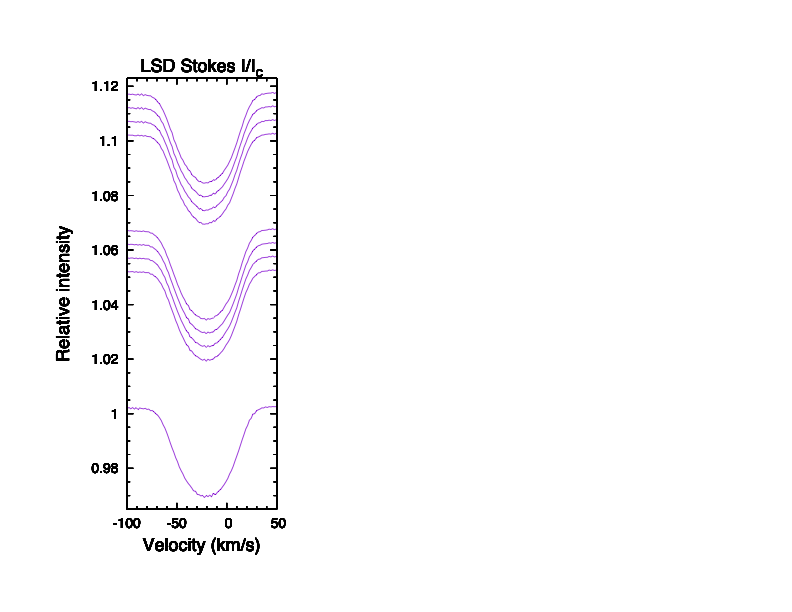} 
   \includegraphics[width=4.7cm, angle=0,trim={1.2cm 0.5cm 16cm 0.8cm}, clip]{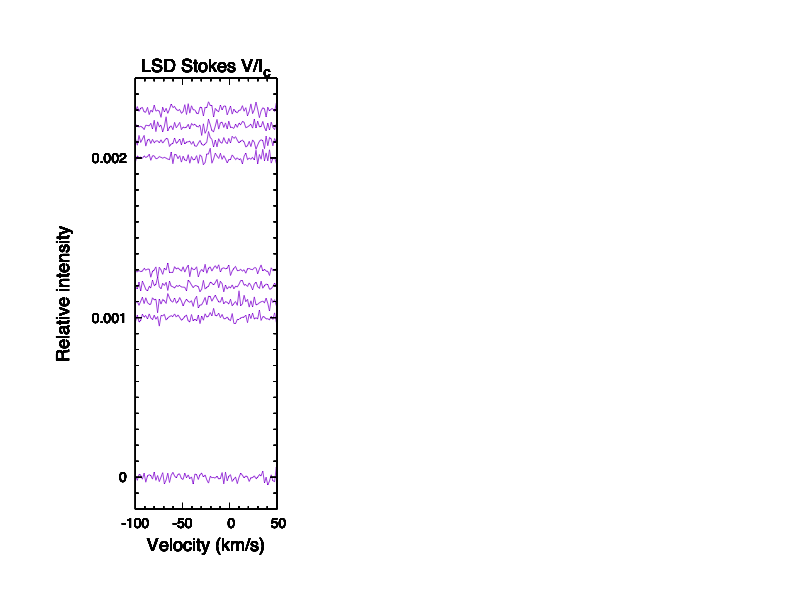}
   \end{tabular}
   \caption{LSD profiles in Stokes I (left panel) and V (right panel) obtained for all acquired spectra of HR~7495 using a mask of 9782 lines. The Stokes V LSD profiles generally show stochastic noise. The spectra are vertically offset by epoch, with the oldest spectrum at the bottom. The corresponding HJD values are listed in Table \ref{table:specpol}, column 1, from top to bottom.
   }
              \label{fig:SpecPol}%
    \end{figure}

The Least-Squares Deconvolution (LSD) technique \citep{Donati97}, as implemented by Colin Folsom  \citep{Erba2024, Erba+24} in Python\footnote{{\rm https://github.com/folsomcp/specpolFlow, https://github.com/folsomcp/LSDpy}}, was employed to probe the mean longitudinal magnetic field ($\langle B_{z}\rangle$). Using the list of atomic data extracted from VALD3 \citep{Piskunov+95, Kupka+99, Kupka+00} for spectral lines visible in a star with $T_{\rm eff}$ =6610~K, log($g$)=3.67 and [M/H]=0.15 we have created a line-mask with 9782 lines to derive LSD profiles for Stokes V and I spectra (see Fig.~\ref{fig:SpecPol}), as well as the LSD profiles for diagnostic null N spectra. 
The last one is used to check the contribution of instrumental polarisation leaked into the studied Stokes V LSD profile. If there is no leaking of polarised signal, it represents non-polarised noise. Based on the derived Stokes V and I LSD profiles, one can measure the mean longitudinal magnetic field and its error bar according to the approach described in detail by \citet{Donati97} and \citet{Kochukhov+2010}. The first, second and third columns in Table~\ref{table:specpol} present, respectively, the Heliocentric Julian Date of the observations, the measured $\langle B_{z}\rangle$, and radial velocities with the corresponding error bars derived from the analysis of Stokes I LSD profiles \citep{Erba+24}. The exposure time is 252 s, except for the first measurement, which is 250 s. The SNR measured at echelle order \#34 in the obtained Stokes I and V spectra is listed in the fourth column.

Our target was observed across three distinct epochs (see Table~\ref{table:data} for details). Analysis of the first spectrum (HJD=2460136.10812) having a high SNR (reaching 820 in Stokes I and 790 in Stokes V spectra at the order \#34 with the central wavelength of $6660$~\AA, see Table~\ref{table:specpol}) has not resulted in a detection of a statistically significant $\langle B_{z}\rangle$. To enhance SNR and reduce the error margins for each of the following two epochs of observations at varied rotational phases (see SNR in Table~\ref{table:data}), we collected four consecutive spectra, each comprising four polarisation scans, striving for the highest SNR within a practical total accumulation time. Unfortunately, none of the observations yielded a statistically significant magnetic field detection in HR~7495 (see Fig.~\ref{fig:SpecPol}). 
The combination of all four obtained spectra results in a negligible value of $\langle B_{z}\rangle$ that is comparable to the derived error bar.
Even when combining two spectra with the largest fields, we obtained only $\langle B_{z}\rangle =-3\pm 2$ G, indicating no significant magnetic field presence at the observed epochs. 
The inspection of the LSD Stokes V profile revealed no discernible signal (see Fig. \ref{fig:SpecPol}), supporting the absence of a significant magnetic field. 
 Employing the false alarm probability (FAP; \citet{Donati+92}), we find the lowest value obtained is 0.025 (for $B_z = -0.2 \pm 3.0$ G, HJD=2460243.71444). For a magnetic detection to be statistically significant, the FAP must be smaller than 0.00001, while a marginal detection requires an FAP between $10^{-3}$ and $10^{-5}$ \citep{Donati97, Hahlin2021}. Based on these criteria, we do not have even a marginal detection of a magnetic field in HR 7495.

It is important to note that the observational scheduling did not allow selecting specific rotational phases optimal for magnetic field detection. Ideally, observations should coincide with the peak visibility of the stellar spots one might infer from the light curve, assuming that its variations are indeed due to magnetism. However, given the low amplitude of the rotational modulation observed, strong magnetic fields were not anticipated. In addition, the complexity of a dynamo-driven magnetic field configuration could also lead to cancellation effects, thus potentially explaining the non-detection of a statistically significant $\langle B_{z}\rangle$. Additional observations should be performed to systematically sample the rotational phase for conclusive results. Unfortunately, our spectropolarimetric observations were not obtained simultaneously with any TESS data.

\begin{table}[h]
\caption{Summary of spectropolarimetric observations of HR~7495 with ESPaDOnS. The exposure time for each observation is 252 seconds, except for the first one, which has an exposure time of 250 seconds.}
\label{table:specpol}
\centering
\begin{tabular}{l c c c}
\hline\hline
HJD  & $\langle B_{z}\rangle \pm \sigma_{B}$ & RV$\pm \sigma_{RV}$ & SNR \\ 
     &   [G]  &  [km s$^{-1}$] & Stokes I/V \\
\hline

2460136.10812 &  2.8 $\pm$ 3.2 & -20.9 $\pm$ 0.1 &  820/790  \\ %2884282
2460238.84517 &  2.0 $\pm$ 3.1 & -20.7 $\pm$ 0.1 &  860/800  \\ % 2918702
2460238.84919 & -0.9 $\pm$ 3.1 & -20.7 $\pm$ 0.1 &  850/790   \\ % 2918706
2460238.85322 & -3.6 $\pm$ 3.1 & -20.7 $\pm$ 0.1 &  850/800  \\ %2918710
2460238.85725 &  0.9 $\pm$ 3.1 & -20.7 $\pm$ 0.1 &  860/810   \\ %2918710
2460243.71444 & -0.2 $\pm$ 3.0 & -20.9 $\pm$ 0.1 &  860/810   \\ % 2919837
2460243.71846 & -2.0 $\pm$ 3.0 & -20.9 $\pm$ 0.1 &  850/810   \\ % 2919841
2460243.72247 & -3.3 $\pm$ 3.1 & -20.9 $\pm$ 0.1 &  850/810   \\ % 2919845
2460243.72649 &  1.3 $\pm$ 3.1 & -20.9 $\pm$ 0.1 &  850/790   \\ % 2919849

\hline
\hline
\end{tabular}
\end{table}

%trim={left bottom right top}

\section{Data analyses}\label{sec:analyses}

\begin{figure}
   \centering
 \includegraphics[width=\columnwidth, trim=0 0 0 70,clip]{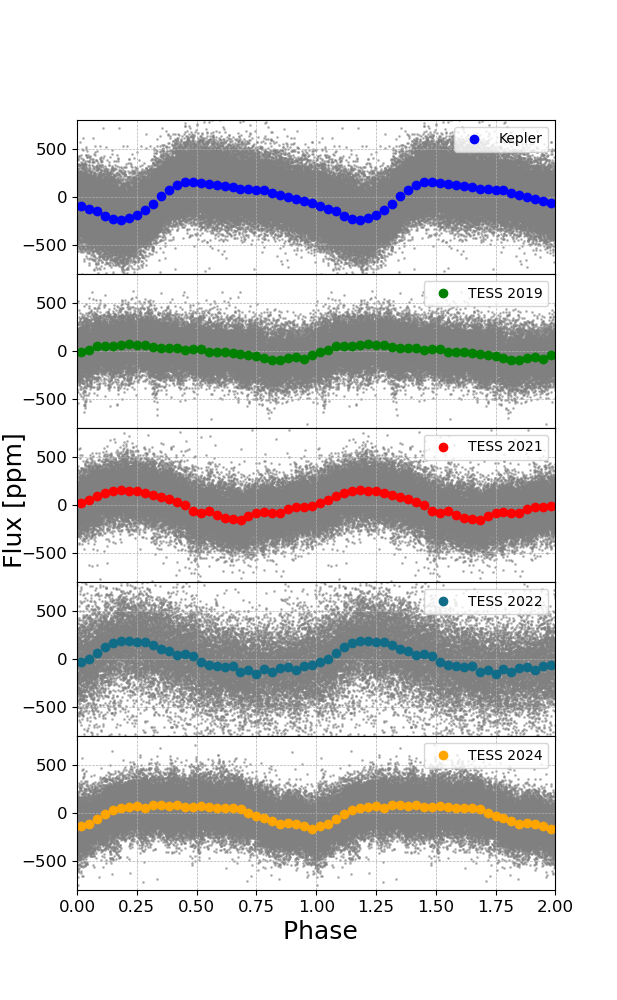}
   \caption{Phase-folded original \textit{Kepler} and TESS data displayed in grey, using the spike frequency derived from \textit{Kepler} data. Coloured curves represent the binned data, showing the median value in each of the 80 arbitrarily chosen bins. }
              \label{fig:phases_all_originial_data}%
    \end{figure}

The \textit{Kepler} mission provided near-continuous observations of HR~7495 for over four years, in contrast to TESS, which collected data in intervals of 27 or 54 days, separated by gaps of 1-2 years (see Table~\ref{table:data} for details). Consequently, we analysed each TESS dataset independently. Using the methods from \citet{Henriksen2023a}, we determined the spike amplitude and frequency across the different TESS observations, as outlined in Table~\ref{table:spike}. These results align with the spike frequency of $0.6770 \pm 0.0005$ cd$^{-1}$ derived from \textit{Kepler} data (see also Table \ref{table:HR7495_params}), which serves as the basis for all analyses. Figure \ref{fig:phases_all_originial_data} displays the photometric data phase-folded with this spike frequency.
In our phase-folded light curve analyses, we aligned all observations to a reference epoch of 2459797.10309 (T0 of the TESS 2022 dataset), ensuring dataset consistency and facilitating comparison of contemporaneous TESS and RV data, as discussed in Section \ref{sec:resulst_discussions}.

\begin{table}
\caption{Average spike amplitudes and frequencies for TESS observations from different years.}
\centering
\label{table:spike}
\begin{tabular}{l c c}        % 3 columns: Year, Frequency, Amplitude
\hline\hline
TESS data set & Frequency [cd$^{-1}$] & Amplitude [ppm] \\
\hline
TESS 2019 & $0.677 \pm 0.016$ & $66 \pm 6$ \\
TESS 2021 & $0.677 \pm 0.009$ & $132 \pm 5$ \\
TESS 2022 & $0.676 \pm 0.020$ & $152 \pm 9$ \\
TESS 2024 & $0.677 \pm 0.010$ & $107 \pm 8$ \\
\hline\hline
\end{tabular}
\end{table}

\begin{figure}
   \centering
 \includegraphics[width=\columnwidth]{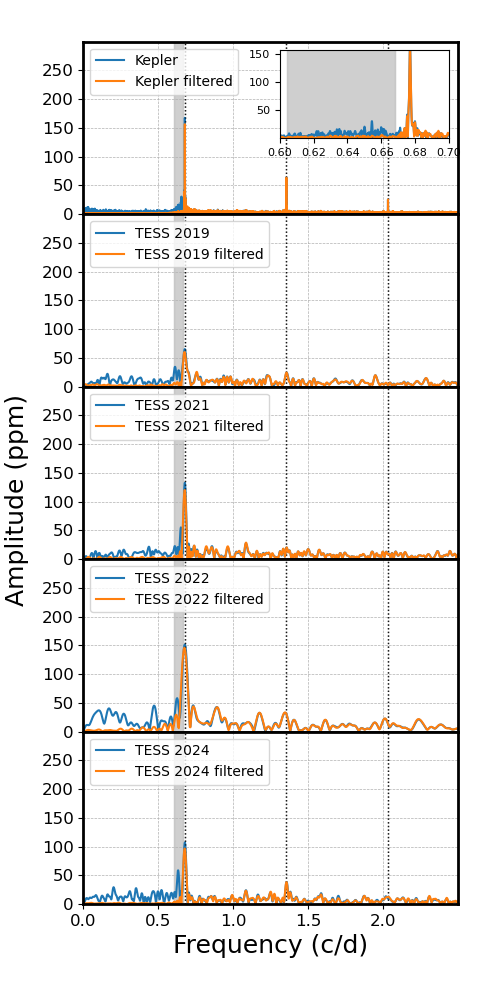}

   \caption{Fourier spectra of \textit{Kepler} and TESS data, with \textit{Kepler} shown in the upper panel and TESS data in panels 2 to 5. Original unfiltered data are depicted in blue, while bandpass-filtered data are shown in orange. The grey area highlights the position of the hump, and dotted black vertical lines mark the spike frequency and its harmonics. No additional signal is found beyond 2.5 cd$^{-1}$.} 
              \label{fig:fourier}%
    \end{figure}

To visualise and interpret the rotational modulation signal (i.e., the spike), we employed a bandpass filter using Weighted Least Squares (details available in Rasmus Handberg's Github repository\footnote{\url{https://github.com/rhandberg/timeseries/tree/master}}). We focused on signals within the frequency range of 0.672 to 5 cd$^{-1}$, although no significant signal was detected above 2.5 cd$^{-1}$. The lower limit was set to exclude the hump component while preserving the harmonics, which are attributed to either the non-sinusoidal shape of the light curve or OsC modes of higher azimuthal order. Figure \ref{fig:fourier} shows the Fourier spectra for all photometric datasets, with the expected hump region shaded in grey. Due to its relatively low amplitude, the hump is primarily discernible in the \textit{Kepler} data (see inset). We display the original (unfiltered) data in blue and the bandpass-filtered data in orange, with dashed vertical lines marking the identified spike frequency and its first and second harmonics. It is noteworthy that the resolution of TESS data is insufficient to separate the hump from the spike signal fully. Nonetheless, the 17 quarters of \textit{Kepler} observations have allowed us to precisely establish the frequency boundaries, with the hump set between 0.604 and 0.668 cd$^{-1}$ (see Table~\ref{table:HR7495_params}). This range is well separated from the bandpass lower boundary at 0.672 cd$^{-1}$ in \textit{Kepler} data, with a difference roughly ten times larger than the \textit{Kepler} resolution of 0.00068 cd$^{-1}$.

\begin{figure*}
   \centering
 \includegraphics[width=\textwidth]{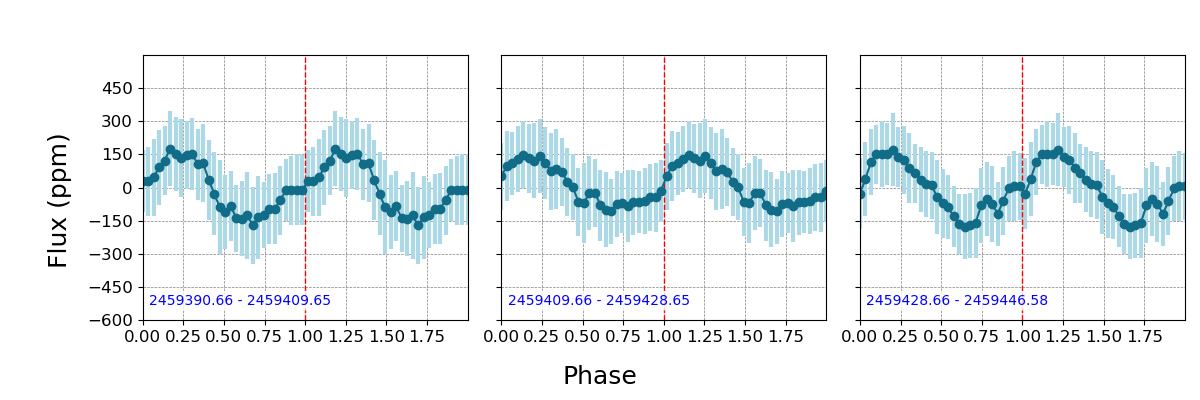}
   \caption{Phase-folded original TESS data. Each panel presents 19 days combined and binned into 30 equal segments. The petrol-blue data points are the binned values with the lighter blue bars representing the standard deviation.  }
              \label{fig:TESS2022_phase_original}%
    \end{figure*}

\begin{figure*}
   \centering
 \includegraphics[width=\textwidth]{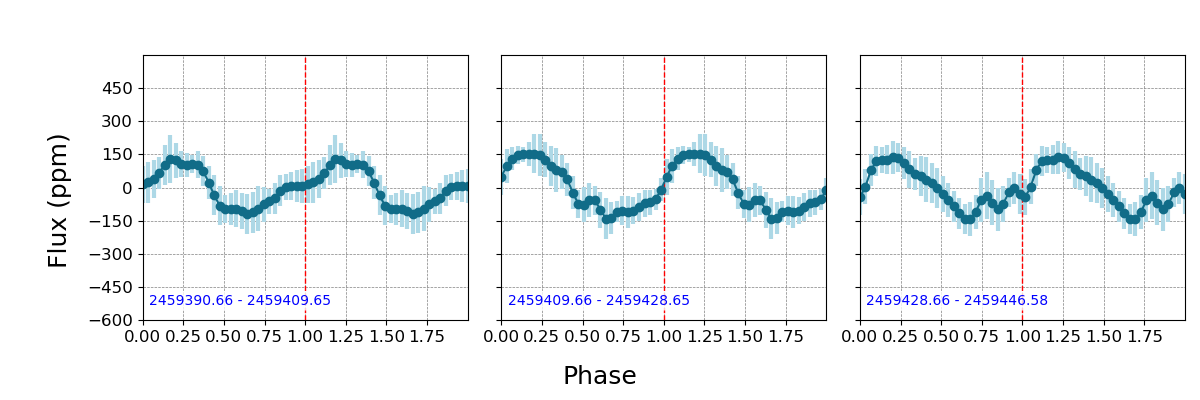}
   \caption{Same as Fig.~\ref{fig:TESS2022_phase_original}, but for bandpass filtered data.}
              \label{fig:TESS2022_phase_bandpass}%
    \end{figure*}

This approach enables a detailed examination of temporal spike variability and potential tracking of spot evolution. Figures \ref{fig:TESS2022_phase_original} and \ref{fig:TESS2022_phase_bandpass} show the original and bandpass-filtered data from the TESS 2021 dataset, illustrating how bandpass filtering reduces scatter in the binned data without significantly altering the overall shape or temporal variability, enhancing our interpretation of the spike signal.

The data were analysed on three timescales to capture different aspects of the spike variability: first, across the entire dataset (Figures \ref{fig:phases_all_originial_data} and \ref{fig:fourier}), allowing determination of average spike parameters (see Tables \ref{table:HR7495_params} and \ref{table:spike}) and enabling comparisons with other stars as in \citet{Henriksen2023a, Henriksen2023b}. The difference in amplitude between the \textit{Kepler} and TESS data is influenced not only by the temporal variability of the spike but also by the difference in bandpass, with TESS being significantly redder than \textit{Kepler}. Second, as shown in Figs.~\ref{fig:TESS2022_phase_original} and \ref{fig:TESS2022_phase_bandpass}, in 19-day windows, corresponding to the \textit{Kepler} spike lifetime (see Table \ref{table:HR7495_params}). The spike lifetime was estimated using the procedure introduced by \citet{Giles2017}, initially for F to M spectral types and later applied by \citet{Henriksen2023a} for hump \& spike stars (see their equation 2.3). This approach calculates the autocorrelation function of the bandpass filtered time series and then fits an exponential model via least-squares minimisation to estimate spike lifetime. The model incorporates parameters such as the decay timescale of the spike,  and the rotation period, capturing periodic modulations from stellar rotation. Third, data were analysed in 150-day segments, allowing high-precision phase determination of the three peaks (the spike and its harmonics) and comparison with theoretical expectations for OsC modes (see Section \ref{sec:theoOsC}).

\begin{figure*}
    \centering
    \includegraphics[width=9cm]{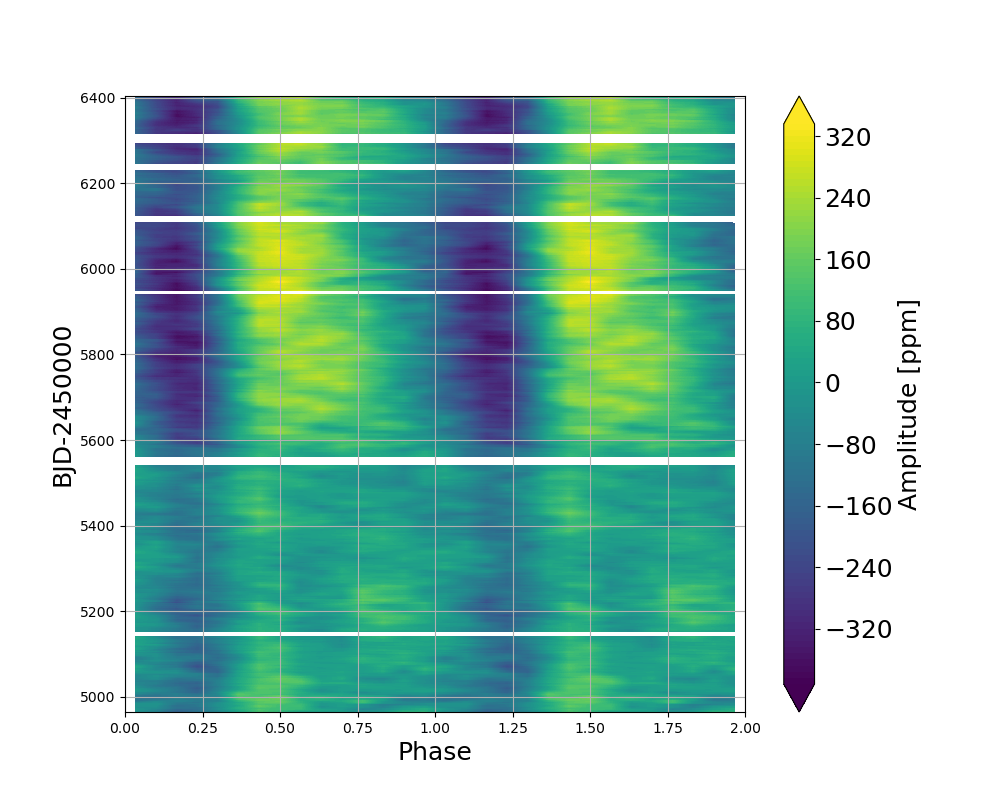}
    \includegraphics[width=9cm]{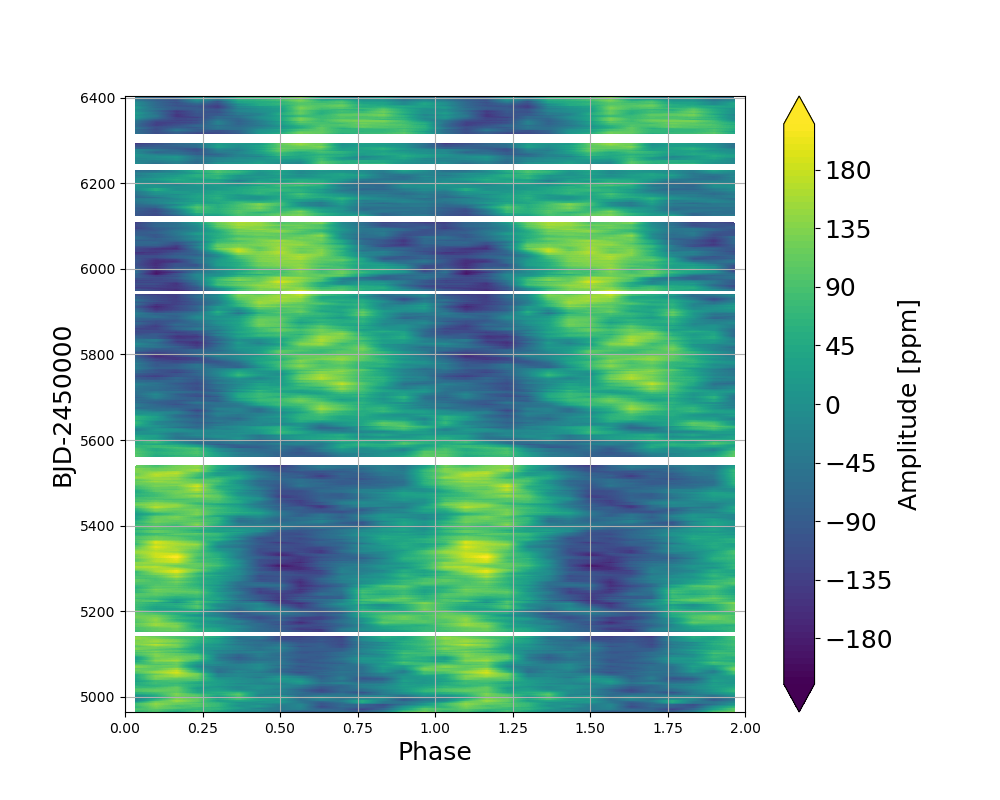}

    \caption{\textit{Kepler} data: Time-dependent phase plot showing the dynamic changes in the phase-folded light curve. The analysis window is 19 days long, shifted by steps of 0.2 days. White areas indicate regions where the data coverage in a given window is less than 70\%. The left panel displays the phase-folded light curve as a function of time, revealing temporal variations. In the right panel, we subtract the average phase-folded light curve (computed from the entire \textit{Kepler} data set) from each windowed phase-folded light curve to highlight the temporal variability. See the text for additional details.  } 
    \label{fig:dynamic_phase}
    
\end{figure*}

\begin{figure*}
   \centering
 \includegraphics[width=\textwidth, trim=0 0 100 10,clip]{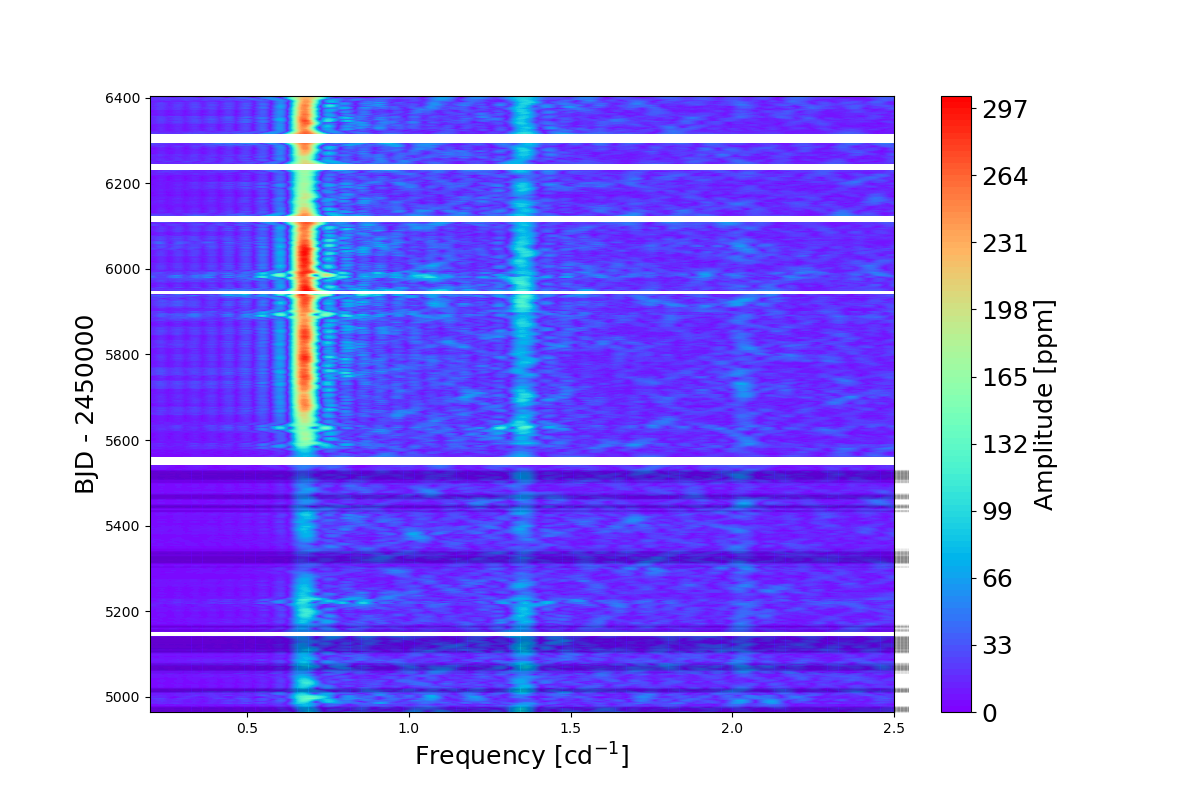}
   \caption{ Time-Resolved Lomb-Scargle Analysis of the \textit{Kepler} data. A sliding window of 19 days was used, with each window shifted by a step size of 0.2 days. The horizontal black lines and the ticks on the right side of the diagram indicate time steps where the amplitude of the first harmonic exceeds that of the fundamental, supporting the interpretation of stellar spots. No significant signal is observed beyond 2.5~cd$^{-1}$.} 

   \label{fig:phase_1stharmonic}%
    \end{figure*}

To investigate the temporal behaviour of the light curve, we examined short-term changes in the phase-folded light curve, shown in the left panel of Fig. \ref{fig:dynamic_phase}. Each line in this figure represents a 19-day window, shifted incrementally by 0.2 days to capture finer variations. This window length, chosen to match the spike lifetime, reveals that while the minimum remains relatively stable, the overall shape of the phase-folded light curve changes significantly. This variability is also evident in the power spectrum, shown in Fig. \ref{fig:phase_1stharmonic}, where each power spectrum was computed for 19-day windows, similarly shifted by 0.2 days. White gaps in both figures indicate data gaps. We required each window to have at least 70\%  data coverage to minimise aliasing effects. During the first 600 days of \textit{Kepler} data, the phase-folded light curve exhibits a double-hump feature consistent with stellar spot behaviour. This feature is also visible in the frequency domain, where the first harmonic of the spike frequency shows a higher amplitude than the spike itself. These intervals are marked with horizontal black dotted lines in Fig. \ref{fig:phase_1stharmonic}. This feature discussed further in Section \ref{sec:resulst_discussions}, strongly indicates the presence of spots.

\begin{figure*}
    \centering
    \includegraphics[width=15cm]{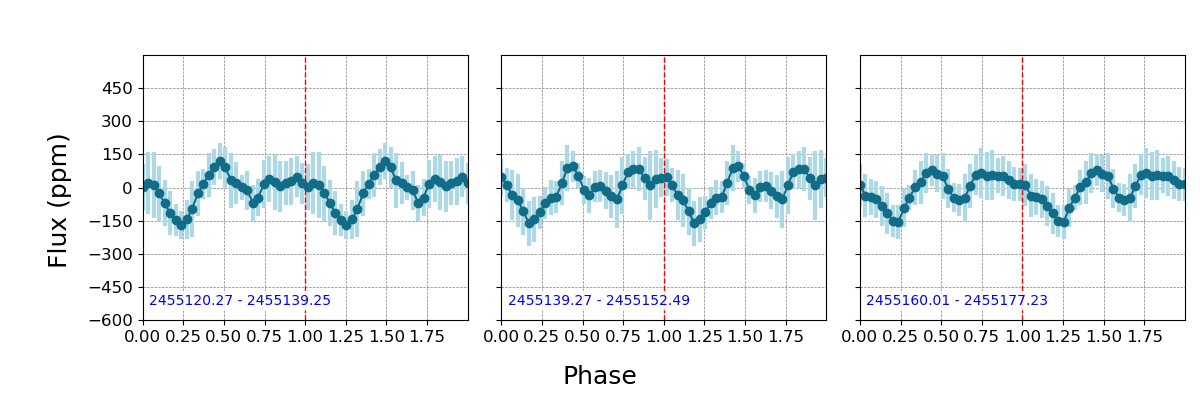}
    \includegraphics[width=15cm]{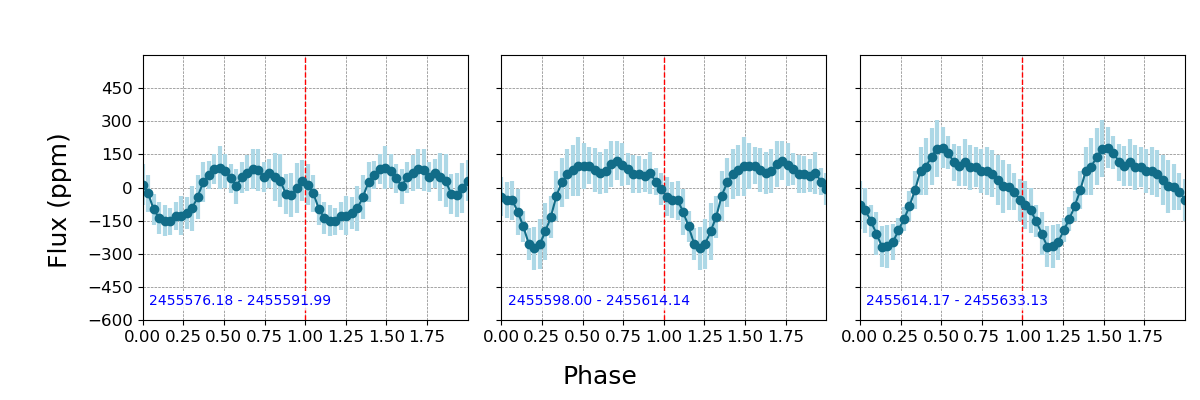}
    \includegraphics[width=15cm]{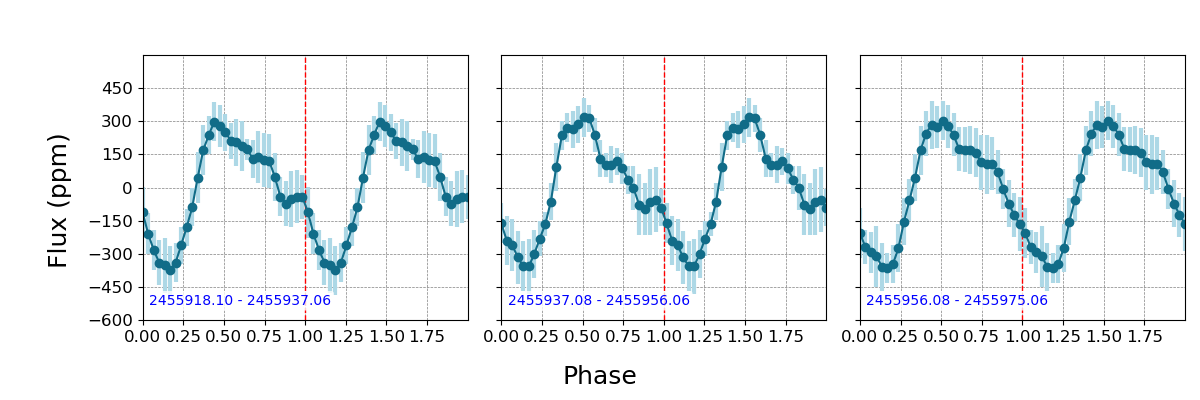}
    \includegraphics[width=15cm]{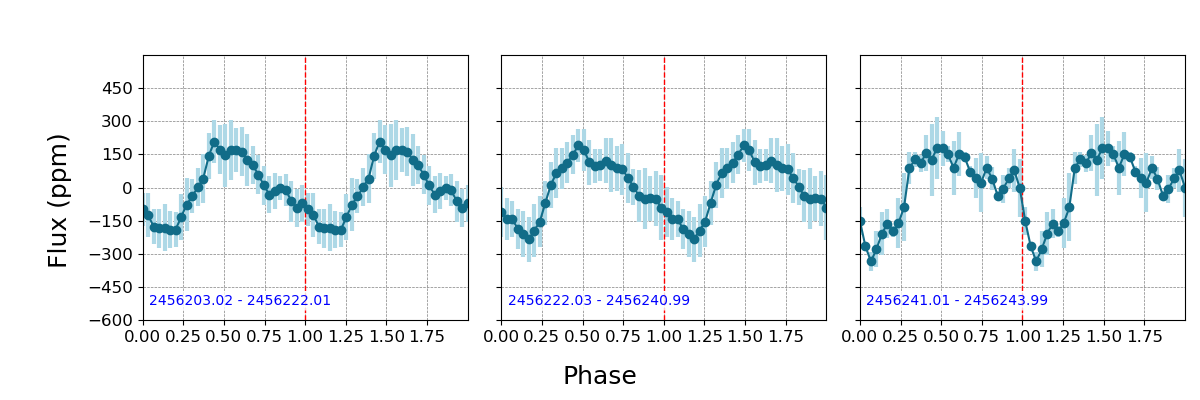}
    
    \caption{\textit{Kepler} data: selected subsets. Each panel shows data combined over a 19-day interval and binned into 30 equal segments. The data in each row are nearly continuous, with the exception of a 7–8 day gap between the second and third panels in the first row and between the first and second panels in the second row. Petrol-blue points represent the binned data, with error bars indicating the standard deviation within each bin.
    } 
    \label{fig:Kepler_phases_3rot}
\end{figure*}

To highlight deviations in the phase-folded \textit{Kepler} light curve from its average behaviour, the right panel of Fig. \ref{fig:dynamic_phase} shows data with the average light curve (upper panel, Fig. 2) subtracted. Since HR~7495 exhibits increased amplitude after approximately BJD 2455600, the average is influenced by this more active phase. Subtracting it reveals clear differences, including changes in the light curve shape and shifts in the phase of minimum flux. This is also reflected in Fig. \ref{fig:Kepler_phases_3rot}, which presents a more traditional view of the phase-folded light curves across four segments of \textit{Kepler} data after bandpass filtering, spaced across the full \textit{Kepler} observation period. The panels span almost consecutive 19-day intervals (listed in each panel), with a 7–8 day gap between the second and third panels in the first row and the first and second panels in the second row. The minimum flux varies by approximately 0.2-0.25 in phase (about 0.3 days), consistent with what is shown in the left panel of Fig. \ref{fig:dynamic_phase}.

\begin{figure}
    \centering
    \includegraphics[width=9.5cm]{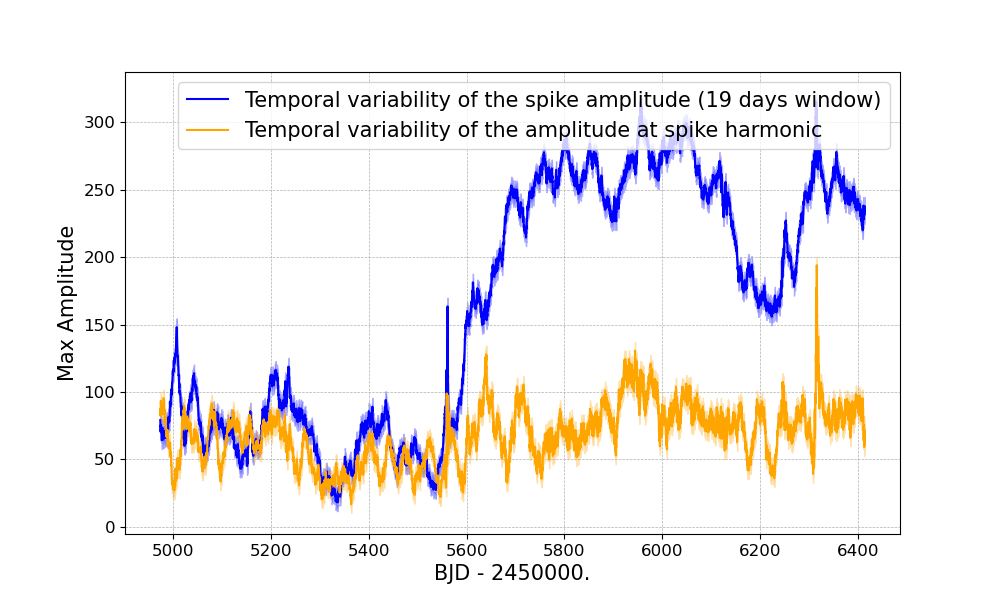}

    \caption{ Amplitude variability of the spike and its harmonic (\textit{Kepler} data). Amplitudes of the spike frequency (blue) and its first harmonic (orange), determined from 19-day windows based on the time-dependent amplitude spectrum shown in Fig. \ref{fig:phase_1stharmonic}. The more transparent blue and orange bands represent the analytical uncertainties in amplitude, respectively.}  
    \label{fig:temp_ampli}
    
\end{figure}

In constructing the amplitude spectrum shown in Fig. \ref{fig:phase_1stharmonic}, we recorded the amplitude in each window for both the spike frequency and its harmonic. To address the lower resolution of the 19-day window, we determined each amplitude by measuring the maximum peak within the range of the spike frequency $\pm$ the window resolution (1/19 days). Figure \ref{fig:temp_ampli} displays these results, highlighting where the first harmonic amplitude exceeds that of the spike frequency. The more transparent blue and orange bands represent the analytical uncertainty in amplitude \citep{MontgomeryODonoghue1999}.

\begin{figure}
    \centering
    \includegraphics[width=\columnwidth]{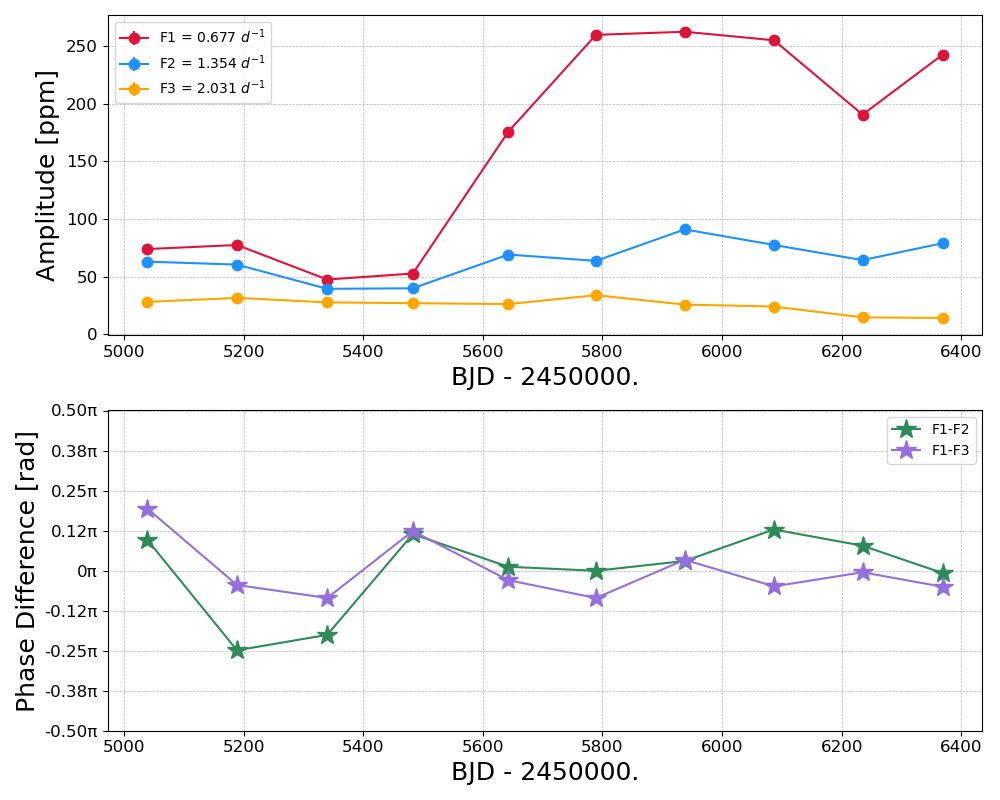}

    \caption{Upper panel: Amplitudes determined for subsets of the \textit{Kepler} data spanning 150 days. The BJD corresponds to the average value for each data bin. Uncertainties are smaller than the symbols depicted. Lower Panel: Phase differences are shown, and the average difference is removed to enhance readability.   
    }
    \label{fig:ampli_phaseshift}
\end{figure}

To compare the data with the model describing OsC modes in Section \ref{sec:theoOsC}, we analysed both the amplitude and, primarily, the phases of the spike by subdividing the light curve into 150-day segments to assess the phase stability/variability across all three peaks. To ensure a sufficiently high signal-to-noise ratio, more extended subsets were necessary. Using Period04 \citep{Period04}, we conducted least-squares fitting, holding the frequencies fixed while allowing the amplitudes and phases of all three peaks to vary. The results, shown in Fig. \ref{fig:ampli_phaseshift}, indicate that variability is driven mainly by amplitude changes in the spike frequency, with minimal changes in the harmonics or phase shifts. The stability in phase shifts aligns with the interpretation of stellar spots, as discussed in more detail in Sections \ref{sec:theoOsC} and \ref{sec:resulst_discussions}. Notably, the amplitude variability in these longer subsets is consistent with measurements in Fig. \ref{fig:temp_ampli}.

%--------------------------------------------------------------------
\section{Theoretical expectations for stellar spots} \label{sec:theospots}%

Here, we look at the possible change in stellar luminosity induced by the presence of surface magnetic fields. Stable, large-scale magnetic fields in thermal equilibrium are expected to lead to very small perturbations to the stellar luminosity. \citet{Fuller2023} predict that even for the strongest observed fields in main sequence stars ($\sim 10$ kG), the bolometric luminosity variation is $\delta L/L \le 10^{-6}$, which at present is not detectable even with high-quality space-based photometry.  

The influence of magnetic fields on systems outside of thermal equilibrium can significantly differ. In stars exhibiting transient magnetic phenomena such as magnetic spots due to dynamo action, the lifespan of these magnetic spots may be substantially shorter than the local thermal timescale of the star, depending on how deep the spots are. 
In stars with convective envelopes transporting a large fraction of the stellar flux, such magnetic fields can temporarily interrupt convective energy transport. This explains why magnetic spots on the Sun appear dark: they are not in thermal equilibrium with the layers below them. In stars with radiative envelopes, magnetic spots are expected to appear bright due to the local decrease in total pressure, which results from the reduction in gas and radiation pressure \citep{Cantiello_hotspots_2011, Cantiello_2019_dynamo}. Predictions for the bolometric fluctuations produced by such magnetic spots depend on their size and the field amplitude. However, such predictions do not consider the effect of thermal diffusion, which might significantly mitigate the flux perturbations, particularly for spots that persist longer than the local thermal timescale \citep{Fuller2023}.

We follow \citet{Cantiello_hotspots_2011} to calculate the bolometric luminosity perturbation induced by a spot at the surface of stars with envelopes dominated by radiative energy transport, keeping in mind values recovered using this theory likely correspond to upper limits. Similarly to \citet{Henriksen2023a}, we write the luminosity perturbation as:

\begin{equation}
\centering
\frac{\Delta L}{L} \simeq \frac{4 \Delta T}{T} = \frac{4\nabla_{\rm rad}}{\beta}.
\label{eq:Lum_ratio}
\end{equation}
 where $\Delta$ represent the deviation from the equilibrium value, $L$ is  the luminosity, $T$ the temperature, $\nabla_{\rm rad}$ the radiative temperature gradient and with

\begin{equation*}
\beta = \frac{P_{\rm tot}}{P_{\rm mag}} = \frac{P_{\rm tot}}{B^2/8\pi},
\end{equation*}
$P_{\rm mag}$ is the component of the pressure due to the presence of the magnetic field, $P_{\rm tot}$ is the total pressure, and $B$ is the magnetic field strength.
This equation provides the luminosity contrast of a magnetic spot compared to the unperturbed stellar luminosity. To compare to observations, one has to account for the relative size of the spot compared to the stellar radius. To do so, we add the filling factor  $f=(r/R)^2$, where $r$ is the radius of the spot, and $R$ is the stellar radius. 
\begin{equation}
\frac{\Delta L}{L} =  \frac{4 \nabla_{\rm rad}}{\beta} \left(\frac{r}{R}\right)^2.
\label{eq:Lum_ratio_geom}
\end{equation}

 Since we are merely trying to establish if magnetic spots could explain the amplitude of the observed luminosity fluctuations, we neglect second-order effects like geometric projection and limb darkening.

 It is important to recognise the interplay between the size of the spot and its temperature (or luminosity) contrast: a large spot with minimal luminosity contrast can result in a brightness change similar to that of a smaller spot with a greater luminosity contrast.
  Using the definition of $\beta$, equation \ref{eq:Lum_ratio_geom} can be written as:
\begin{equation}
\frac{\Delta L}{L} =  \frac{4 \nabla_{\rm rad} B^2}{8 \pi P_{\rm tot} } \left(\frac{r}{R}\right)^2
\label{eq:Lum_ratio_geom3}
\end{equation}
 We can then replace the ratio $\Delta L/L$ with the observed photometric amplitude ($A_{\rm spike}$) in the \textit{Kepler} and TESS data. Note that our calculation assumes a single spot, and we do not correct for the instrument bandpass. 

 Since we have limits on the observed magnetic field amplitude, we can then calculate the minimum size of a single  magnetic spot able to induce the observed luminosity variation:
\begin{equation}
    f \simeq  \frac{A_{\rm spike}}{B^2} \frac{ 2 \pi P_{\rm tot} }{\nabla_{\rm rad}} 
    \label{eq:filling}
\end{equation}
Using the stellar models described in \citet{Henriksen2023a} we recover values of $\nabla_{\rm rad}$ and  $P_{\rm tot}$ suitable for HR~7495 ($\nabla_{\rm rad}\approx 0.13$ and $P_{\rm tot}\approx 8\times10^3$  barye). Plugging in 15~G as the upper limit for the magnetic field (see table~\ref{table:specpol}) and 100 ppm as typical spike amplitude, this implies a minimum filling factor $f \approx 0.17$, or a spot minimum radius $r\approx 0.42\, R$.

 We emphasise that our approach considers a highly simplified scenario with a large poloidal field, which could produce a spot with a filling factor, $f$. This setup could account for the observed photometric variability with a filling factor of 0.17 and a magnetic field strength $B$ near the detection limit. While spots could undoubtedly be smaller and have a stronger $B$-field that would remain undetected, this exercise shows that a poloidal $B$-field below the detection threshold could plausibly be responsible for the observed photometric variability.
Indeed, the situation is likely more complex, and figures \ref{fig:TESS2022_phase_bandpass}, \ref{fig:Kepler_phases_3rot}, and \ref{fig:phot_RV_Mag} suggest that several magnetic spots may be present on the surface of HR~7495, complicating the interpretation of the light curve and the derivation of the filling factor. This complexity also makes the adopted limit on the maximum magnetic field associated with surface spots uncertain, as cancellation effects likely influence spectropolarimetric detection.

\section{Theoretical expectation of the phase-folded light curve  in the context of OsC modes} \label{sec:theoOsC}%

In the context of OsC modes, the three peaks observed in the Fourier transforms in  Fig.~\ref{fig:fourier} correspond to prograde sectoral g modes with azimuthal orders of $m=1, 2, 3$. We adopt the convention that $m>0$ indicates prograde modes. 
If the convective core rotates slightly faster than the envelope,
high radial order g modes could be resonantly excited by the OsC modes in the convective core \citep{LeeSaio2020, Lee2021}.
Writing the rotation rates of the core and envelope as $\nu^{\rm co}_{\rm rot}$ and $\nu^{\rm en}_{\rm rot}$, respectively, we have a relation
$$\Delta\nu_{\rm rot}\equiv \nu^{\rm co}_{\rm rot} - \nu^{\rm en}_{\rm rot} \ll \nu^{\rm co}_{\rm rot} ~,\nu^{\rm en}_{\rm rot}.$$
As the envelope (rotating at $\nu^{\rm en}_{\rm rot}$) experiences OsC frequencies of  $m\Delta\nu_{\rm rot}$, high radial order g modes  are expected to be resonantly excited with frequencies,
$\nu_{\rm cor}^{m}\sim m\Delta\nu_{\rm rot}$.

In the inertial frame, these g modes have frequencies of

$$\nu_{\rm int}^m = \nu_{\rm cor}^m + m\nu^{\rm en}_{\rm rot} \approx m\nu^{\rm en}_{\rm rot}$$

 where $\nu_{\rm cor}$ refers to the frequency in the corotating frame, and $\nu_{\rm int}$ refers to the frequency in the inertial frame, as observed. Thus, these prograde g modes with $m=1,2,3$ can be identified with the three spikes seen in the Fourier diagram for the \textit{Kepler} data in Figure\,\ref{fig:fourier}.

\begin{figure}
\centering
   \includegraphics[width=\columnwidth]{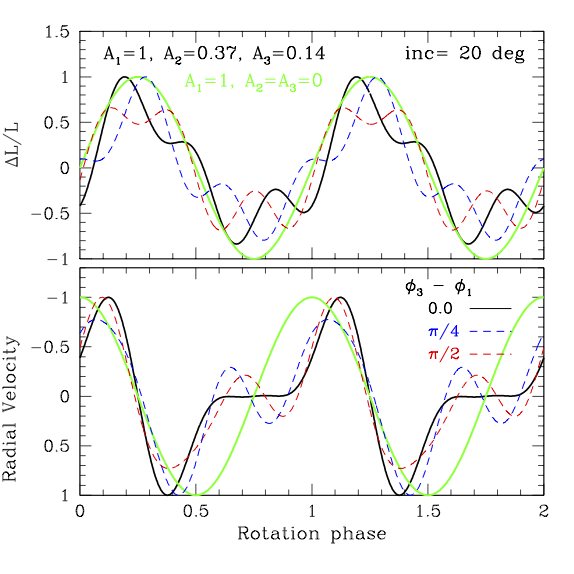}

   \caption{Theoretical predictions for OsC modes comparing light and RV variations, normalized so that their maxima are unity. Upper panel: Light curves predicted for the superposition of sectoral prograde g modes with $m=1, 2, 3$, using amplitude ratios of $1:0.37:0.14$, consistent with the amplitudes in the \textit{Kepler} Fourier diagram (Fig.~\ref{fig:fourier}). The black solid line represents a model where the initial phases of all modes are identical. To investigate the impact of phase shifts, the $m=3$ mode is adjusted as shown, while the $m=2$ mode’s initial phase is aligned with that of the $m=1$ mode. The green curve illustrates a scenario where only the spike ($m=1$) is observable, with no contributions $m=2$ and $m=3$ modes. Lower panel: RV variations predicted for the same prograde g modes, using the same amplitude ratios and phase offsets as in the upper panel. The green line corresponds to a model with only the $m=1$ mode. See text for more details.}
    \label{fig:theoretical_light_RV_curves}
    \end{figure}

An OsC mode in the convective core couples with a prograde low-frequency g mode in the envelope, whose temperature and velocity variations on the surface should be observed as brightness and radial-velocity variations. 
The angular dependence of temperature variations and velocity of low-frequency g modes in a rotating star is represented by the Hough function and its derivatives with respect to $\cos \theta$ as discussed in, e.g., \citet{LeeSaio1997} and \citet{Townsend2003} in detail. We have used formulae of \citet{LeeSaio1997} to calculate temperature-perturbation and velocity distribution on the stellar surface as functions of time. Then, assuming an inclination angle of the line of sight to the rotation axis, we integrated the brightness and radial-velocity distributions on the visible stellar surface at each rotation phase.
We have added results for each mode of $m=1,2,3$ adopting a set of weights, given by the average amplitudes in the Kepler data $(A_1, A_2, A_3)$ and highlighted in Fig. \ref{fig:theoretical_light_RV_curves}. The resulting light and radial-velocity variations as a function of time are plotted in Fig. \ref{fig:theoretical_light_RV_curves} as a function of the rotation phase.

The frequency of a resonantly coupled g mode, $\nu_{\rm cor}$, in the corotating frame of the envelope is much smaller than the envelope rotation frequency $\nu^{\rm en}_{\rm rot}$ so that the spin parameter of the mode is very high $2\nu^{\rm en}_{\rm rot}/\nu_{\rm cor}\gg 1$.
Among the g modes with high spin parameters, only prograde sectoral modes are visible, having broadly distributed amplitudes on the surface, symmetric to the equator, while the other g modes are confined into a narrow zone around the equator and anti-symmetric to the equator \citep[see e.g. Appendix in][]{Saio2018a}. 
For this reason, we have computed a synthetic light curve in Figure\,\ref{fig:theoretical_light_RV_curves} by superposing the prograde sectoral g modes of $m=1,2,3$ with amplitude ratios of $1:0.37:0.14$ that are consistent with the amplitudes in the \textit{Kepler} Fourier diagram shown in Figure\,\ref{fig:fourier} (top panel).

 While the Hough function is dependent on the spin parameter, it becomes nearly independent for prograde sectoral modes when the spin parameter is sufficiently large. We assumed a high spin parameter of $2\nu^{en}_{rot} /\nu^m_{cor} = 30/m$ for each g mode with azimuthal order $m$, corresponding to an assumed small difference between the rotation frequencies of the convective core and the envelope. Additionally, considering a limb-darkening parameter of $\mu=0.6$ and an inclination of the line-of-sight to the rotation axis, we integrated the local flux perturbation (assumed to be proportional to the temperature perturbation) on the visual surface at each pulsation phase.

In Figure\,\ref{fig:theoretical_light_RV_curves}, we have adopted an inclination angle of $20^\circ$; the results remain insensitive to this angle because the amplitudes of prograde g modes are broadly distributed in latitude. Each curve displayed represents simulated light or RV variations, incorporating the \textit{Kepler} amplitude ratios that contribute to the bumps observed in the light curve shown in this figure.
These bumps arise from the contributions of the $m=2$ and $3$ modes, and their locations depend on the phase relationships among the $m=2$ and 3 modes. In Figure\,\ref{fig:theoretical_light_RV_curves}, we illustrate the effects of phase shifts in the $m=3$ mode. The black solid line represents the case with no initial phase shift, while the red and blue dashed lines correspond to phase shifts of  $\pi/4$ and $\pi/2$, respectively. As expected, these phase shifts alter the bump locations in the light curve. While the frequencies of the three modes in the corotating frame differ only slightly, these minor differences ($|\nu^{\rm m}_{cor}-\nu^{\rm m'}_{cor}|$) can slowly modify the phase relationships among the g-modes. This gradual evolution may explain the observed slow variations in the folded light curves. In the case of the RV curve (lower panel, Fig. \ref{fig:theoretical_light_RV_curves}), the phase shifts variations have minimal impact on the location of the maximum and minimum RV shift. The RV `shoulder' showing no variability between phases 0.5 and 1 is caused by the $m=1, 2$ and $3$ modes cancelling out their effects. In green we show the model for only one $m=1$ mode. It is interesting to note that, in all cases, the zero RV shift closely aligns with the maxima and minima in the light variation. The relationship between the light and RV maxima/minima, along with the shape of the phase-folded curves, will serve as an essential tool for studying other hump \& spike stars. We refer to the next section for a comparison between this model and our observations (Section \ref{sec:resulst_discussions}).

\section{Results and Discussion}\label{sec:resulst_discussions}

This study aims to differentiate between two scenarios: (1) rotational modulation caused by co-rotating spots driven by magnetic fields and (2) oscillation modes in the convective core that resonantly excite g-modes reaching the surface. \citet{Henriksen2023a} conducted ensemble analyses of all \textit{Kepler} hump and spike stars, concluding that scenario (1) is more likely; however, scenario (2) could not be entirely ruled out. Currently, HR~7495 stands out as the most promising target for follow-up and in-depth analyses.

\subsection{Stellar spots scenario}
In the context of stellar spots, the clear photometric variations observed in a star with a low inclination of approximately 20$^\circ$ suggest that the spots responsible are not circumpolar. The flux changes (see Fig.~\ref{fig:TESS2022_phase_bandpass} and Fig.~\ref{fig:Kepler_phases_3rot}) imply that these spots are located at lower latitudes, allowing them to rotate in and out of view at least partially, despite the nearly pole-on perspective. 

The phase-folded data shown in Fig.~\ref{fig:TESS2022_phase_bandpass} and Fig.~\ref{fig:Kepler_phases_3rot} exhibit evolving features consistent with evolving spots. This variability is also apparent in Fig.~\ref{fig:dynamic_phase}, which depicts changes in the entire phase-folded \textit{Kepler} light curve over timescales aligned with the spike's lifetime. The minimum flux varies between phases 0 and 0.25, corresponding to a shift of about 0.3 days, as illustrated in Fig.~\ref{fig:dynamic_phase}. This quarter-phase shift may be attributed to star spots migrating in latitude over time; however, we currently lack a quantitative dynamo model that would allow us to predict the latitudes where spots will appear or their migration patterns. In the case of the Sun, spots move from higher to lower latitudes as part of the activity cycle, influenced by latitudinal differential rotation and meridional circulation. The effects of these processes on stellar spots in stars like HR~7495, which has a shallow convective envelope, remain an open question. However, it is possible that the driving forces for meridional circulation are weaker in stars with shallow convective envelopes, likely resulting in reduced circulation strength. This, in turn, suggests that spot migration may also be limited.

The \textit{Kepler} data clearly show an increase in spike amplitude in the second half of the mission (see Fig. \ref{fig:temp_ampli} and Fig.~\ref{fig:ampli_phaseshift}). In the magnetic spot scenario, this trend suggests the presence of a dynamo that could generate activity cycles similar to those observed in other stars \citep[e.g.,][]{Jeffers2023}. There is a significant decrease in amplitude around BJD 2456200, but the amplitude rises again toward the end of the \textit{Kepler} dataset. Conducting ensemble analyses of other stars with similar hump \& spike features may provide further insights into this phenomenon and allow for comparisons with low-mass stars, where dynamo activity is well established.

In the case of stellar spots, spike harmonics in the Fourier spectrum emerge due to the non-sinusoidal shape of the light curve and can exhibit amplitudes higher than that of the fundamental spike frequency for spot distributions giving rise to strongly non-sinusoidal variation. For OsC modes, the harmonics correspond to modes with different azimuthal orders, and their amplitudes are expected to decrease monotonically due to geometrical effects. \citet{Henriksen2023a} demonstrated that out of 162 stars analysed from the entire \textit{Kepler} dataset, 36 showed harmonics consistent with stellar spot interpretations. For HR~7495, as illustrated in Fig. \ref{fig:fourier}, none of the spike harmonics exceed the spike amplitude. However, segmenting the \textit{Kepler} dataset revealed that at certain times, the amplitude of the first harmonic can exceed the spike amplitude, as shown by the horizontal lines in Fig. \ref{fig:phase_1stharmonic}. This behaviour results from a double-wave phase-folded light curve, as further demonstrated in the lower section of Fig. \ref{fig:dynamic_phase} (left panel) and strongly suggests that the spike in HR~7495 is associated with stellar spots, confirming that the spike frequency corresponds to the surface rotation frequency.

We collected semi-contemporaneous TESS and RV data, as shown in Fig. \ref{fig:phot_RV_Mag}, to compare with the model presented in Section \ref{sec:theoOsC}. The upper panel displays the phase-folded, bandpassed TESS data set from 2022, while the lower panel presents the phase-folded RV curve. The RV data observed simultaneously with the photometric data and the data collected 19 days prior (spike lifetime) are highlighted in red. The data within 30 days of the TESS observations are shown in blue. The slight discrepancy in timing is due to the scheduled TESS observations from Sector 54 not being delivered correctly for reasons that remain unknown. 

Figure \ref{fig:phot_RV_Mag} illustrates that both the flux and RV curves exhibit non-sinusoidal shapes. Notably, the flux maxima and minima correspond with zero radial velocity shifts, while the maximum blue shifts occur just before the flux minima. This pattern suggests the presence of one or more spots on the stellar surface. However, the variability observed in Figures \ref{fig:dynamic_phase} and \ref{fig:phase_1stharmonic} indicates that it is more likely that multiple spots are involved.

A similar pattern was observed in HD~188774 \citep{Lampens2013}, where the maximum blue shift also preceded the minimum flux. In contrast to HR~7495, HD~188774 displays a double-wave pattern, implying the presence of two opposing spots. Furthermore, HD~188774 has a detected magnetic field below 100 G, as reported by \citet{Neiner&Lampens2015}.

One challenge in interpreting these spots is the observed RV amplitude, which is at least an order of magnitude larger than expected if spots were solely responsible for the photometric variability. This unexpectedly large RV amplitude in HR~7495 mirrors findings for HD~188774, where \citet{Lampens2013} also reported disproportionately large variations in RV relative to flux. The discrepancy is unlikely due to the redder TESS bandpass, as HD~188774 shows similar issues when comparing \textit{Kepler} data with RV measurements. Conducting simultaneous observations of HR~7495 in multiple bandpasses (e.g., TESS and Plato) along with concurrent RV measurements could help clarify these effects.

The relative amplitudes of photometric and RV variations depend on the number, size, overall coverage (filling factor), and distribution of the spots. Degeneracies can arise because different configurations of spots -- varying in size, number, and latitude -- can produce similar signals. Observed phase offsets between photometric and RV signals provide valuable insights into the properties of the spots. However, fully disentangling these overlapping effects will require additional data and a more detailed analysis.

\subsection{OsC modes scenario}
\begin{figure*}
   \centering
 \includegraphics[width=\textwidth]{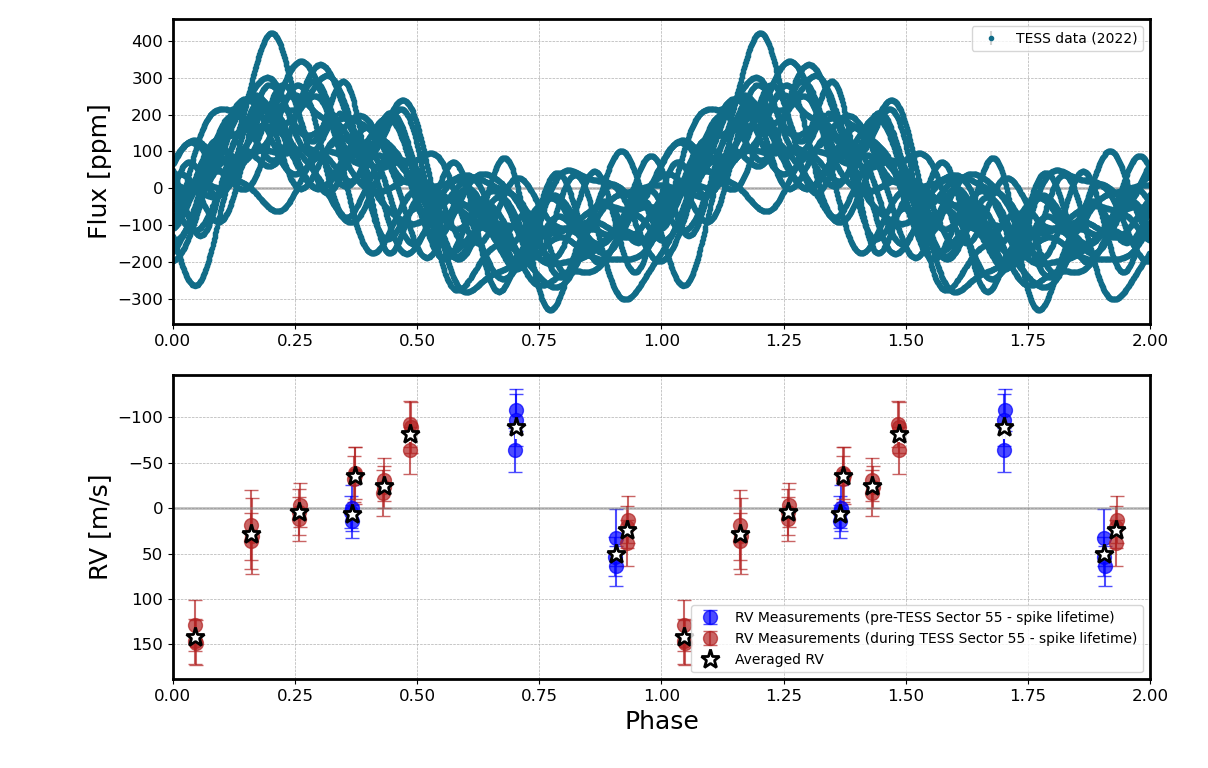}

   \caption{Upper panel: Phase-folded bandpass-filtered data TESS 2022 data. Lower panel: Phase-folded RV data observed simultaneously or just before the TESS 2022 (Sector 55) data. The red data points represent individual observations during the TESS observing period up to 19 days prior, which reflects the spike lifetime, while the blue data points go back 30 days in time. The stars indicate the average value for each bin. Note that the RV measurements for a given epoch are obtained from consecutive spectra (see Table \ref{table:rv_data}). }
\label{fig:phot_RV_Mag}
    \end{figure*}

In the context of OsC modes, a comparison between Fig. \ref{fig:theoretical_light_RV_curves} and Fig. \ref{fig:phot_RV_Mag} indicates that the model does not replicate the contemporaneous observations accurately. In Fig. \ref{fig:fourier} (panel 4), it is evident that the first harmonic is not statistically significant due to its very low amplitude, while the second harmonic is completely unobservable. This implies that we would only have one mode, and the model curve shown in green in Fig. \ref{fig:theoretical_light_RV_curves} is appropriate for comparison with our observations. Regarding RV variability, the presence of only one pulsation mode would also correspond to the model shown in green, resulting in a maximum blue shift at phase 0. Comparing the observations with the model shows that the observed RVs are shifted by 0.5 in phase, where the maximum blue shift is around phase 0.5. Therefore, we conclude once again that the observed spike is caused by stellar spots rather than an oscillation (OsC) mode.

While we lack RV measurements from the \textit{Kepler} observing period, we can still compare our findings with the light variations shown in Fig.~\ref{fig:theoretical_light_RV_curves} (upper panel) to investigate whether the temporal variability of the peaks is due to phase changes in the $m=2$ and $m=3$ modes. If the spike resulted from OsC modes, its finite lifetime could be related to the convective turnover timescale in the core, as discussed by \citet{Henriksen2023a}. However, the transformation of the light curve from a double peak to a more single-wave shape in the initial \textit{Kepler} data (see Fig.~\ref{fig:dynamic_phase}) would suggest significant phase shifts between the pulsation modes. As shown in Fig.~\ref{fig:ampli_phaseshift}, the phase differences between the primary spike (F1) and its first (F2) and second (F3) harmonics start at approximately $\pi/2$ rad (0.25 cycle) but stabilise to less than $\pi/4$ (0.125) for most of the dataset. These phase differences are critical because they reveal relationships among the phases of individual peaks; significant phase shifts would distort the phase-folded light curve, resembling more the red dashed curve in Fig.~\ref{fig:theoretical_light_RV_curves}. We observe notable amplitude variability throughout the dataset, including the disappearance of the second harmonic peak ($m=3$, assuming OsC modes) as illustrated in Fig.~\ref{fig:ampli_phaseshift}. Based on Fig.~\ref{fig:ampli_phaseshift}, we conclude that this variability in the \textit{Kepler} data is primarily driven by amplitude changes rather than phase shifts, which aligns with the phase-folded light curve in Fig.~\ref{fig:dynamic_phase}. Consistently, the minimum flux occurs between phases 0 and 0.25 cycles (0 to $\pi/4$ rad) across the \textit{Kepler} observations (see Fig.~\ref{fig:Kepler_phases_3rot}). 
The relatively stable phases throughout the dataset suggest that, if OsC modes were present, we would expect light curve variations similar to those depicted by the black or blue curves, with harmonic phase shifts below $\pi/4$ rad (refer to Fig.~\ref{fig:theoretical_light_RV_curves}). However, the considerable shape variations shown in Figs.~\ref{fig:ampli_phaseshift} and \ref{fig:Kepler_phases_3rot} imply that small phase shifts alone cannot account for the observed differences.

\par

\subsubsection{Evolutionary stage of HR~7495}

\begin{figure*}
\centering
\includegraphics[width=\columnwidth]{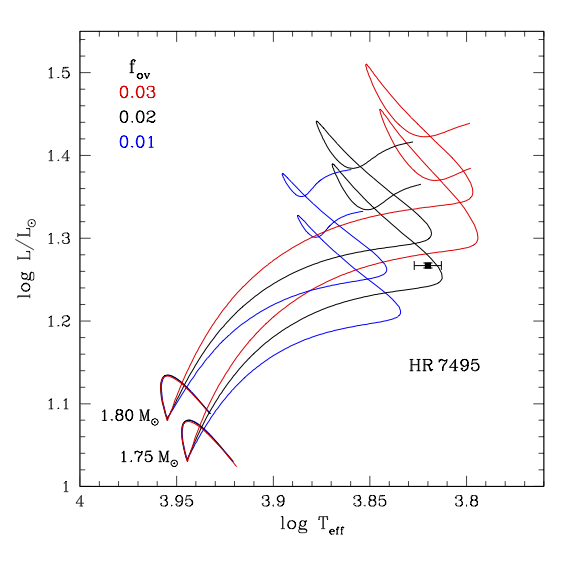}
\includegraphics[width=\columnwidth]{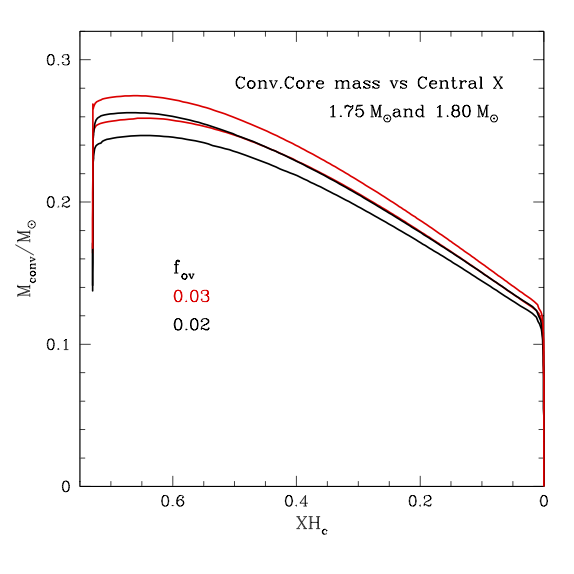}
\caption{Left Panel: Evolutionary tracks for masses ${\rm M}=1.75, 1.80~{\rm M}_\odot$ with initial chemical composition $(X,Z)=(0.73,0.01)$. The adopted parameters of diffusive overshooting at the convective core boundary, $f_{\rm ov}$, are colour-coded as indicated in the upper left corner. The position of HR~7495 is marked by a filled square with error bars. Right panel: Convective core mass versus central hydrogen abundance X, with $f_{\rm ov}$ color-coded.}
\label{fig:evol}%
\end{figure*}

Another aspect that can help distinguish between the two scenarios is the evolutionary stage of HR~7495. OsC modes can only be expected if the stellar core is both convective and sufficiently large. Figure \ref{fig:evol} illustrates the main-sequence evolution of stars with masses of $1.75$ and $1.80~{\rm M}_\odot$ and an initial chemical composition of $(X,Z)=(0.73,0.01)$. These simulations were performed using MESA v.7184 \citep{pax11,pax13,pax15}, with the same settings as in \citet{Saio2018a}; i.e. we employed the Schwarzschild criterion for the convective/radiative boundary, included elemental diffusion to smooth the Brunt-Väisälä frequency, and implemented radiative turbulence to prevent excessive helium settling in the outer envelope. Rotation was not included in the analysis as its impact on evolutionary models and their positions on the Hertzsprung-Russell Diagram (HRD) is relatively minor for qualitative discussions. Instead, the analysis focuses on the more significant parameter related to the extent of core overshooting. The position of HR~7495 on the HRD is indicated by a black square, with parameters from Table~\ref{table:HR7495_params}. It is evident that HR~7495 is a relatively evolved star, which may still possess a convective core. Depending on the overshooting parameter, $f_{\rm ov}$, it could be close to the terminal age main sequence (TAMS) or beyond.
Stars with the masses specified above have a small convective core ranging from $0.25$ to $0.1~{\rm M}_\odot$ during their main-sequence evolution, as depicted in the right panel of Figure \ref{fig:evol}. As hydrogen is consumed in the core, the star becomes more luminous and cooler in the HR diagram. The evolutionary track turns abruptly to the left when the hydrogen abundance in the convective core decreases to less than $0.05$, at which point the convective core disappears. Therefore, for OsC modes to be excited and explain the observed spikes, the diffusive overshooting parameter, $f_{\rm ov}$, should be greater than $0.02$, and the mass ${\rm M}$ should be at least $1.75~{\rm M}_\odot$, given the adopted initial chemical composition. Since we do not have additional constraints for HR~7495, we cannot conclusively determine which $f_{\rm ov}$ may be appropriate. However, comparing to $\gamma$ Doradus stars, which have similar masses, \citet{Mombarg2021} showed that stars with masses around $1.80~{\rm M}_\odot$ tend to have a rather large $f_{\rm ov}$, which would suggest that HR~7495 is near the end but still on the main sequence.

%--------------------------------------------------------------------

\section{Conclusions}\label{sec:conclusions}

HR~7495, the brightest `hump \& spike' star known with a visual magnitude of 5.06, is an ideal target to determine whether the spike is of magnetic origin, induced by spots on the stellar surface, or corresponds to OsC modes that propagate to the surface by resonantly exciting g modes. Our target was observed with \textit{Kepler} for 4.5 years and with TESS during four different seasons. Additionally, we gathered RV measurements contemporaneously with the TESS 2022 data and obtained spectropolarimetric data from three different epochs.

Analysing the shape of the phase-folded light curves, we conclude that they are more consistent with stellar spots based on their shapes and temporal evolution, which suggests evolving stellar spots. We also noticed that several subsets of the \textit{Kepler} data show the first harmonic peak of the spike having a higher amplitude than the main spike (see Fig.~\ref{fig:phase_1stharmonic}). This pattern is unexpected for oscillation modes but suggests the presence of stellar spots.

 The \textit{Kepler} data reveal a significant increase in spike amplitude during the latter half of the mission. This suggests the possibility of dynamo-driven activity cycles, with a brief dip in amplitude around BJD 2456200 before increasing again. Comparative analyses with other stars exhibiting similar characteristics could improve our understanding and provide valuable comparisons with low-mass stars with well-documented dynamo activity.

 Our simultaneous 2022 TESS and RV observations show non-sinusoidal shapes in both flux and RV curves, with flux maxima and minima coinciding with zero radial velocity shifts and maximum blue shifts occurring just before flux minima (Fig. \ref{fig:phot_RV_Mag}). This behaviour, also observed in the magnetic $\delta$ Scuti star HD~188774 \citep{Lampens2013, Neiner&Lampens2015}, along with the temporal variability in Fig.~\ref{fig:dynamic_phase}, suggests that HR~7495 has multiple spots, consistent with a dynamo-generated magnetic field. 

We found no significant magnetic field, which is not unexpected given the overall low spike amplitude and the evidence for multiple stellar spots. While contemporaneous photometric measurements are lacking, the \textit{Kepler} and TESS data strongly suggest the existence of multiple persistent spots. Therefore, we can reasonably anticipate that this behaviour remains consistent during our spectropolarimetric observation period. This inferred spot multiplicity could complicate the detection of a significant magnetic field using the LSD method due to potential cancellation effects. Meanwhile, additional phase-resolved spectropolarimetric observations of HR~7495 with higher SNR are needed to check for line profile variability and signatures of magnetic fields.

 In the context of OsC modes, we developed a theoretical model to investigate the overall shape, temporal variability, and phase relationships in phase-folded light and RV curves. Our findings indicate that the observed phase shift between contemporaneous light and RV measurements (Fig.~\ref{fig:phot_RV_Mag}) is half a phase different from what our model predicts (Fig.~\ref{fig:theoretical_light_RV_curves}). Furthermore, the phase shifts between various peaks in the \textit{Kepler} data are too small to explain the overall temporal variability in the shape of the phase-folded light curve.

While HR~7495 may still be at the end of its main sequence, potentially possessing a convective core large enough to excite OsC modes, the evidence strongly supports a different explanation. The observed spike is consistent with surface rotational modulation induced by stellar spots, suggesting a weak magnetic field. This conclusion is further reinforced by the findings of \citet{Henriksen2023a}, which show that 36 out of 162 stars also favour the stellar spots interpretation. Given this comprehensive body of evidence, it is prudent to move beyond the OsC scenario in explaining the spikes observed in all `hump \& spike' stars.

As demonstrated by \cite{Henriksen2023b}, there is a discernible correlation between the amplitude of the spike and the power of the Rossby and gravity mode humps, respectively. \footnote{Approximately half of the 'hump \& spike' stars exhibit a hump following the spike, indicative of unresolved high-order gravity modes \citep[for more details, see][]{Henriksen2023b}.} Our conclusion that the spike is linked to magnetic fields enables further exploration of the role of a subsurface convection zone dynamo and the presence of Rossby and gravity modes.
Further, we intend to focus future efforts on characterising the stellar spots of HR~7495 through additional contemporaneous TESS observations and detailed and high-cadence spectroscopy to analyse line profile variations in different elements. Expanding our search in TESS data for more bright `hump \& spike' stars will further our ability to identify more candidates for magnetic field measurements using spectropolarimetry.

\begin{acknowledgements}
Co-funded by the European Union (ERC, MAGNIFY, Project 101126182 ). Views and opinions expressed are, however, those of the authors only and do not necessarily reflect those of the European Union or the European Research Council. Neither the European Union nor the granting authority can be held responsible for them. The Center for Computational Astrophysics at the
Flatiron Institute is supported by the Simons Foundation. We acknowledge \textit{Kepler}, TESS, NOT, CFHT, Python, and the use of ChatGPT4 to assist in refining the grammar and help in writing the plotting procedures. 
This paper includes data collected with the TESS mission, obtained from the MAST data archive at the Space Telescope Science Institute (STScI). Funding for the TESS mission is provided by the NASA Explorer Program. STScI is operated by the Association of Universities for Research in Astronomy, Inc., under NASA contract NAS 5–26555. We thank the referee for their valuable comments and suggestions, which have helped improve this manuscript.

\end{acknowledgements}

\bibliographystyle{aa} 
\bibliography{main_ArXiV}

\begin{appendix} %First appendix
\section{Radial velocity measurements with FIES.}
\begin{table}[h]
\caption{Radial velocity observations of HR 7495. The SNR is the averaged signal-to-noise ratio computed from the red and the blue part of the spectrum. The spectrum highlighted with the asterisk is the template we used to extract the RV measurements. }
\label{table:rv_data}
\centering
\begin{tabular}{cccc}
\hline\hline
BJD & RV [ms$^{-1}$] & RV$_{err}$ [ms$^{-1}$] & SNR \\
\hline
2459767.4210620 & 53.16 & 21.43 & 155 \\
2459767.4225670 & 32.62 & 31.51 & 147 \\
2459767.4240100 & 63.86 & 22.28 & 149 \\
2459772.5318530 & 4.02 & 17.61 & 186 \\
2459772.5333270 & 14.31 & 18.03 & 208 \\
*2459772.5348040 & 0 & 25.37 & 211 \\
2459777.4590920 & -63.82 & 24.07 & 189 \\
2459777.4605710 & -97.47 & 28.58 & 188 \\
2459777.4620350 & -108.35 & 23.49 & 188 \\
2459787.4792280 & -92.8 & 25.26 & 154 \\
2459787.4807010 & -63.93 & 26.01 & 160 \\
2459787.4821640 & -89.03 & 28.32 & 168 \\
2459791.5759690 & 11.8 & 24.57 & 154 \\
2459791.5774470 & 4.27 & 25.09 & 146 \\
2459791.5789000 & -3.81 & 24.43 & 134 \\
2459798.4758930 & 38.57 & 25.54 & 180 \\
2459798.4773910 & 21.07 & 23.01 & 170 \\
2459798.4788420 & 12.49 & 25.57 & 177 \\
2459800.6054390 & -32.51 & 25.39 & 183 \\
2459800.6069090 & -38.52 & 28.63 & 186 \\
2459800.6083960 & -36.21 & 31.47 & 188 \\
2459807.5096020 & 146.55 & 24.99 & 163 \\
2459807.5111030 & 128.82 & 27.92 & 173 \\
2459807.5125530 & 148.38 & 24.77 & 170 \\
2459813.5872670 & 18.78 & 38.7 & 75 \\
2459813.5887580 & 36.01 & 30.54 & 95 \\
2459813.5902660 & 30.53 & 41.61 & 109 \\
2459818.4188420 & -16.76 & 25.19 & 164 \\
2459818.4203100 & -31.57 & 23.92 & 171 \\
2459818.4217870 & -23.55 & 23.13 & 160 \\
\hline
\end{tabular}
\end{table}

\end{appendix}

\end{document}